\documentclass[a4paper,twoside,english,british,useAMS,usenatbib]{mn2e}
\usepackage{helvet}
\usepackage[T1]{fontenc}
\usepackage[latin9]{inputenc}
\usepackage{color}
\usepackage{array}
\usepackage{textcomp}
\usepackage{multirow}
\usepackage{amstext}
\usepackage{amssymb}
\usepackage{graphicx}
\usepackage{esint}
\usepackage[authoryear]{natbib}

\makeatletter

\pdfpageheight\paperheight
\pdfpagewidth\paperwidth

\providecommand{\tabularnewline}{\\}

\newenvironment{lyxlist}[1]
{\begin{list}{}
{\settowidth{\labelwidth}{#1}
 \setlength{\leftmargin}{\labelwidth}
 \addtolength{\leftmargin}{\labelsep}
 }}
{\end{list}}

\title[The Arecibo Ultra-Deep Survey]{\textbf{A blind \textsc{H\,i} Mass Function from the Arecibo Ultra-Deep Survey (AUDS)}}\author[L.  Hoppmann et al.]{L.  Hoppmann$^{1,3}$\thanks{E-mail:laura.hoppmann@icrar.org}, L. Staveley-Smith$^{1,2}$,
W. Freudling$^{3}$, M. A. Zwaan$^{3}$,\newauthor  R. F. Minchin$^{4}$, M. R. Calabretta$^{5}$
\\$^{1}$International Centre for Radio Astronomy Research (ICRAR), M468, University of Western Australia, 35 Stirling Hwy, Crawley, WA, 6009, Australia\\$^{2}$ARC Centre of Excellence for All-sky Astrophysics (CAASTRO)
\\$^{3}$European Southern Observatory (ESO), Karl-Schwarzschild-Str. 2, 85748 Garching, Germany
\\$^{4}$Arecibo Observatory, HC03 Box 53995, Arecibo, PR 00612, USA
\\$^{5}$CSIRO Astronomy and Space Science, PO Box 76, Epping, NSW 1710, Australia
}

\usepackage{datetime}
\usepackage{babel}
\usepackage{subfig}
\usepackage{datetime}
\usepackage{multicol}

\newcommand{\nat}{Nature}
\newcommand{\mnras}{Monthly Notices of the Royal Astronomical Society}
\newcommand{\apj}{The Astrophysical Journal}
\newcommand{\apjl}{The Astrophysical Journal Letters}

\newcommand{\aap}{Astronomy and Astrophysics}
\newcommand{\aj}{The Astronomical Journal}

\newcommand{\Sint}{\ensuremath{S_\mathrm{int}}}
\newcommand{\Sints}{\ensuremath{S_\mathrm{int}^*}}
\newcommand{\MHz}{\mathrm{MHz}}

\newcommand{\mJyMHz}{\mathrm{mJy\,MHz}}
\newcommand{\OMHI}{\ensuremath{\Omega_\hi}}

\newcommand{\MHI}{M_\textsc{H\,i}}
 
\newcommand{\hi}{\textsc{H\,i}}

\newcommand{\Vmax}{\ensuremath{V_\mathrm{max}}}
\newcommand{\Vmaxc}{\ensuremath{V_\mathrm{max,CV}}}
\newcommand{\FreudOMHI}{ 3.74\pm 1.70}

\newcommand{\al}{{-1.53\pm  0.04}}
\newcommand{\Mi}{ {9.87\pm0.02}}
\newcommand{\PL}{{3.52\pm 0.80}}

\newcommand{\alC}{{-1.37 \pm 0.03}}
\newcommand{\MiC}{{9.75\pm  0.041}}
\newcommand{\PLC}{{7.72\pm  1.4}} 

\newcommand{\OI}{{1.76\pm 0.09} } %
\newcommand{\OIC}{ 2.22\pm0.10 } %
\newcommand{\OS}{ 1.63\pm 0.08}  %
\newcommand{\OSC}{{2.33\pm  0.07}}  %

\newcommand{\OSSWML}{ 1.94\pm 0.49} %
\newcommand{\OISWML}{ 1.96 \pm 0.51} %
\newcommand{\alSWML}{ -1.14\pm0.10}
\newcommand{\MiSWML}{ 9.70 \pm0.07}
\newcommand{\PLSWML}{ 9.82\pm 5.40}

\newcommand{\RI}{(6.24\pm 0.23)\times 10^{7} \,h\, M_\odot \mathrm{Mpc^{-3}} }
\newcommand{\RIC}{(6.53\pm 0.31)\times 10^{7} \,h \,M_\odot \mathrm{Mpc^{-3}}}

\newcommand{\RS}{(4.46\pm 0.21)\times 10^{7} \,h \,M_\odot \mathrm{Mpc^{-3}}}
\newcommand{\RSC}{(5.78\pm 0.20)\times 10^{7} \,h \,M_\odot \mathrm{Mpc^{-3}}}

\newcommand{\OSZlow}{3.68 \pm 0.39 } %
\newcommand{\OSZhigh}{1.93\pm  0.19 }  
\newcommand{\OSZlowL}{ 5.33 \pm 1.89}  %
\newcommand{\OSZhighH}{ 2.24\pm 0.78 }%
\newcommand{\NNo}{102}
\newcommand{\OSbest}{ 2.63\pm  0.10}

\AtBeginDocument{
  
}

\makeatother

\usepackage{babel}
\begin{document}
\maketitle
\begin{abstract}
The Arecibo Ultra Deep Survey (AUDS) combines the unique sensitivity
of the telescope with the wide field of the Arecibo L-band Feed Array
(ALFA) to directly detect 21~cm $\hi$ emission from galaxies at
distances beyond the local Universe bounded by the lower frequency
limit of ALFA ($z=0.16$). AUDS has collected 700~hr of integration
time in two fields with a combined area of 1.35~deg$^{2}$. In this
paper we present data from 60\% of the total survey, corresponding
to a sensitivity level of $80\,\mathrm{\mu Jy}$. We discuss the data
reduction, the search for galaxies, parametrisation, optical identification
and completeness. We detect $\NNo$ galaxies in the mass range of
$\log(M_{\hi}/M)-2\log h=5.6-10.3$. We compute the $\hi$ mass function
(HIMF) at the highest redshifts so far measured. A fit of a Schechter
function results in $\alpha=\alC$, $\Phi^{*}=(\PLC)\,\times10^{-3}h^{3}\,\mathrm{Mpc^{-3}}$
and $\log\,(M_{\hi}^{*}/M_{\odot})=(\MiC)+2\log h$. Using the measured
HIMF, we find a cosmic $\hi$ density of $\OMHI=(\OSC)\times10^{-4}\, h^{-1}$
for the sample ($z=0.065$). We discuss further uncertainties arising
from cosmic variance. Because of its depth, AUDS is the first survey
that can determine parameters for the $\hi$ mass function in independent
redshift bins from a single homogeneous data set. The results indicate
little evolution of the co-moving mass function and $\OMHI$ within
this redshift range. We calculate a weighted average for $\OMHI$
in the range $0<z<0.2$, combining the results from AUDS as well as
results from other 21~cm surveys and stacking, finding a best combined
estimate of $\OMHI=(\OSbest)\times10^{-4}\, h^{-1}$.
\end{abstract}

\section{Introduction}

Understanding how and at what rate stars form from cool atomic and
molecular gas ($<10^{4}\,\mathrm{K}$) is one of the crucial questions
of modern astrophysics. The star formation rate (SFR) is well measured
from UV, optical, infrared and radio continuum observations, and is
found to increase by an order of magnitude over the redshift interval
of $0<z<2.5$ \citep{Hopkins2006}. By comparison, the evolution of
the atomic and molecular cosmic gas density appears to be less dramatic
with recent galaxy evolution models suggesting that there may only
be a weak evolution of cosmic gas density at $z<2$ if there exists
a self-regulated equilibrium between the inflow of gas into galaxies
and the SFR, \citep{obreschkow2009,Lagos2011,power2010}. However,
better observations are necessary to further develop these models
and to better understand the balance between gas accretion, star formation
and feedback.

Measurements of the $\mathrm{H}_{2}$ density (e.g. \citealt{Keres2003ApJ...582..659K})
are unfortunately not easy, as the molecule does not possess a low-energy
rotational transition. Instead, we are dependent on the use of CO
as a proxy, with its uncertain dependence on optical depth, metallicity
and the radiation field. More accurate observations are available
for atomic hydrogen via Damped Lyman-$\alpha$ (DLA) systems at high
redshift or via the 21~cm line in the local Universe.

DLAs are wide absorption features caused by high column densities
of $\hi$ $(>2\times10^{20}\mathrm{{atoms\, cm}^{-2}}$) normally
associated with galaxies. DLAs can be observed against bright background
sources such as QSOs and appear to represent objects which contain
the majority of $\hi$ at redshifts $1.6<z<5.0.$ \citep{Wolfe2005}
and are therefore an important reservoir for star formation.

DLA measurements at $z\approx4$ find values of $\OMHI$ around double
the local value \citep{Zwaan2005}. \citet{Prochaska2005} and \citet{Prochaska2009}
find a 50\% decrease to occur at lower redshifts $2.3<z<5.5$. However,
Noterdaeme et al. (2012) use a larger Sloan Digital Sky Survey (SDSS)
Data Release (DR) 9 sample and only find a 20\% decrease over a similar
range. Measurements using $\textsc{\ensuremath{\mathrm{Mg}}ii}$ systems
at low-redshift DLA proxies are consistent with the latter \citep{rao2006},
though selection effects are uncertain and errors are high. Nevertheless
evolution in $\Omega_{\hi}$ above redshift $z>0.5$ appears significantly
lower than the corresponding evolution in the star formation rate.

Interpretation of 21~cm observations is more robust, but sensitivity
considerations mean that observations are mainly limited to the local
Universe. Examples of 21~cm surveys are summarised in Table \ref{tab:Selection-of-21cmX}.
Extensive mapping of the sky has been done since the installation
of multi-beam receivers on the Arecibo, Parkes and Effelsberg telescopes,
which transformed these telescopes into powerful survey facilities.
This resulted in a significant increase in the area surveyed and the
number of galaxies detected. In terms of redshift, however, these
surveys are still limited to $z\approx0$. The two largest $\hi$
surveys are the $\hi$ Parkes All-Sky Survey (HIPASS) ($z<0.04$)
\citep{Meyer2004,Wong2006} and the Arecibo Legacy Fast ALFA (ALFALFA)
($z<0.06)$ \citep{Giovanelli2005}. HIPASS detected 5,317 galaxies
in the southern and the northern sky up to a declination of $\delta=+25.5$.
ALFALFA observed an area of $\sim7000\mathrm{\,{deg}^{2}}$ with the
target of detecting about 30,000 galaxies at 21~cm. 

\begin{table*}
\protect\caption{Selection of major blind 21~cm spectral line surveys. \label{tab:Selection-of-21cmX}}

\noindent \centering{}{\footnotesize{}}%
\begin{tabular}{l||llcccc}
\hline 
\multicolumn{2}{l}{{\footnotesize{}Survey}} & \multicolumn{1}{l}{{\footnotesize{}Reference}} & \multicolumn{1}{c}{{\footnotesize{}Telescope}} & \multicolumn{1}{c}{{\footnotesize{}N}} & \multicolumn{1}{c}{{\footnotesize{}Area (deg$^{2}$)}} & \multicolumn{1}{c}{{\footnotesize{}Redshift }}\tabularnewline
\hline 
\hline 
\multicolumn{2}{l}{{\footnotesize{}AHISS$^{a}$ }} & {\footnotesize{}\citet{Zwaan1997} } & {\footnotesize{}Arecibo } & {\footnotesize{}$66$} & {\footnotesize{}65 } & {\footnotesize{}$0-0.025$}\tabularnewline
\multicolumn{2}{l}{{\footnotesize{}ADBS$^{a}$ }} & {\footnotesize{}\citet{Rosenberg2002} } & {\footnotesize{}Arecibo } & {\footnotesize{}$265$} & {\footnotesize{}430 } & {\footnotesize{}$0-0.027$}\tabularnewline
\multicolumn{2}{l}{{\footnotesize{}HIPASS$^{a}$ }} & {\footnotesize{}\citet{Meyer2004}} & {\footnotesize{}Parkes } & {\footnotesize{}$4315$} & {\footnotesize{}21341 } & {\footnotesize{}$0-0.042$}\tabularnewline
\multicolumn{2}{l}{{\footnotesize{}North. HIPASS$^{a}$}} & {\footnotesize{}\citet{Wong2006}} & {\footnotesize{}WSRT} & {\footnotesize{}$1002$} & {\footnotesize{}7997} & {\footnotesize{}$0-0.042$}\tabularnewline
\multicolumn{2}{l}{{\footnotesize{} 40\% ALFALFA$^{a}$}} & {\footnotesize{}\citet{Haynes2011AJ....142..170H}} & {\footnotesize{}Arecibo } & {\footnotesize{}$10,119$} & {\footnotesize{}2799} & {\footnotesize{}$0-0.06$}\tabularnewline
\multicolumn{2}{l}{{\footnotesize{}ALFALFA$^{b}$}} & {\footnotesize{}\citet{Giovanelli2005}} & {\footnotesize{}Arecibo} & \selectlanguage{english}%
{\footnotesize{}$\sim25000$}\selectlanguage{british}%
 & {\footnotesize{}7000 } & {\footnotesize{}$0-0.06$}\tabularnewline
\multicolumn{2}{l}{{\footnotesize{}AGES}\foreignlanguage{english}{{\footnotesize{}$^{b}$
WAPP}}} & {\footnotesize{}\citet{Auld2006}} & {\footnotesize{}Arecibo} & {\footnotesize{}$\sim1300$} & {\footnotesize{}105} & {\footnotesize{}$0-0.06$}\tabularnewline
\multicolumn{2}{l}{{\footnotesize{}~~~~~~~~~~~}\foreignlanguage{english}{{\footnotesize{}Mock}}} &  &  & {\footnotesize{}$\sim1300$} & {\footnotesize{}95} & {\footnotesize{}$0-0.16$}\tabularnewline
\multicolumn{2}{l}{{\footnotesize{}CHILES}\foreignlanguage{english}{{\footnotesize{}$^{b}$}}{\footnotesize{}
prec.}} & {\footnotesize{}\citet{Fernadenz2013}} & {\footnotesize{}VLA} & {\footnotesize{}33} & {\footnotesize{}0.3} & {\footnotesize{}$0-0.193$}\tabularnewline
\multicolumn{2}{l}{{\footnotesize{}ALFA ZOA$^{b}$ shallow}} & {\footnotesize{}\citet{Henning2010AJ....139.2130H}} & {\footnotesize{}Arecibo} & {\footnotesize{}$\sim500$} & {\footnotesize{}1000} & {\footnotesize{}$0-0.06$}\tabularnewline
\multicolumn{2}{l}{{\footnotesize{}~~~~~~~~~~~~~~~~~~ deep}} &  &  & {\footnotesize{}$\sim1500$} & {\footnotesize{}280 } & {\footnotesize{}$0-0.16$}\tabularnewline
\multicolumn{2}{l}{{\footnotesize{}AUDS$^{a}$ prec.}} & {\footnotesize{}\citet{Freudling2011} } & {\footnotesize{}Arecibo } & {\footnotesize{}$18$} & {\footnotesize{}$0.069$} & {\footnotesize{}$0.07-0.16$}\tabularnewline
\hline 
\multicolumn{7}{l}{{\footnotesize{}$^{a}$ Observations completed and survey completely
published.}}\tabularnewline
\multicolumn{7}{l}{{\footnotesize{}$^{b}$ Observations ongoing or data not completely
published - survey results not finalised. }}\tabularnewline
\end{tabular}
\end{table*}

Direct detections of galaxies beyond the local Universe are not only
limited by sensitivity, but also by radio frequency interference (RFI)
and receiver bandwidth. Target galaxies for deep 21~cm observations
are therefore normally preselected. For example, \citet{Catinella2008}
targeted galaxies up to $z=0.25$ to look for the most $\hi$-massive
objects, selecting them by their H-$\alpha$ emission. \citet{Zwaan2001}
and \citet{Verheijen2007} targeted clusters with redshifts about
$z\approx0.2$ to increase the chance of detection.  {Such
a strategy leads to samples that are biased towards galaxies with
high surface brightness in optical bands. Furthermore,} in order to
extend beyond the local Universe, such surveys need to be very sensitive.
With that in mind, we commenced the Arecibo Ultra Deep Survey (AUDS)
- a blind 21~cm survey with the Arecibo L-band Feed Array (ALFA)
to search for 21~cm $\hi$ line emission at redshifts between 0 and
0.16, the limit of the receiver. The AUDS precursor observations \citep{Freudling2011}
were an important test of the feasibility of such a survey. The precursor
survey detected 18 galaxies in the redshift range $0.07<z<0.16$ with
a total integration time of 53~hr. While this provided a measurement
of $\Omega_{\mathrm{\hi}}$ in good agreement with measurements in
the local Universe, the AUDS precursor was limited to a very small
region with few detections and covered only a limited range of redshifts.
Small number statistics and cosmic variance were therefore problems,
resulting in large errorbars.

In this paper, we present results of 60\% of the data from the full
survey. It is a fully sampled and a significantly more sensitive data
set than the precursor observation, providing a larger sample of direct
21~cm detections.  {The total integration time is
eight times larger than that of the precursor survey. This paper allows
a preliminary release for galaxies so far detected, and makes significant
advances in the understanding of the evolution of the $\hi$ mass
function (HIMF). }

Throughout this paper we use $H_{0}\,=\,100h\,\mathrm{km\, s^{-1}\, Mpc}^{-1}$,
$\Omega_{\mathrm{M}}=0.3$ $ $and $\Omega_{\Lambda}=0.7$.

\subsection{Survey Strategy}

\begin{figure}
\centering{}\includegraphics[width=1\columnwidth]{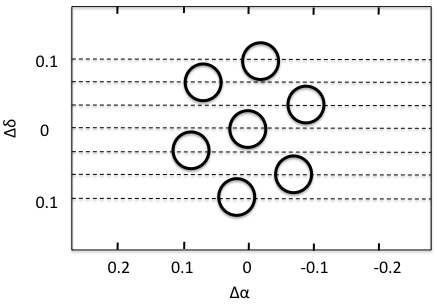}\protect\caption[Idealised relative position of the seven independent ALFA beams.]{\label{fig:beam1} Idealised relative positions of the seven independent
ALFA beams. For AUDS the beams are rotated between $15^{\circ}$ and
$23^{\circ}$ to achieve a uniform coverage of the survey area. }
\end{figure}

The primary goal of AUDS is to systematically survey the cosmic $\hi$
density $(\OMHI)$ in a volume that is beyond the local Universe to
a sensitivity that has not been probed before. The strategy of the
survey is therefore orthogonal to other current single-dish $\hi$
surveys in the sense that AUDS covers a very small area on the sky
using the most sensitive 21~cm system currently available and using
a very long integration time. Our goal was to achieve an exposure
time of about 40~hr per pointing, and a total of 1000~hr of observing
time (including overheads) were assigned to this project. To cover
the field in the most uniform and sensitive way, we used the ``drift
and chase'' mode which we extensively tested and refined during our
precursor observations \citep{Freudling2011}. The basic strategy
was to carry out repeated drift scans over the same field. To optimise
uniformity of the sky coverage, the feed array was rotated to ensure
equal spacing between the beams. Due to the elliptical projection
of the array onto the sky the rotation angle varies between $15^{\circ}$
and $23^{\circ}$. The orientation of the array relative to the scan
direction is shown in Figure \ref{fig:beam1}.

One major difference with the precursor observations was that, with
significantly more observing time available, we were able to Nyquist-sample
the sky using short adjacent drift scans that were offset in declination
by 0.1 of the beam size. This enabled much more accurate determination
of source positions (limited only by signal-to-noise ratio) and source
fluxes.

All observations were carried out at night time to achieve the best
baselines and lowest system temperature. In order to be able to schedule
observations for the project throughout the year, we elected to divide
the total area into two fields at opposite right ascension within
the SDSS footprint. The size of the individual field areas was determined
by science goals and telescope limitations. A small area is required
to go deep. On the other hand, the upper redshift limit of ALFA means
that once sufficient sensitivity to low-mass galaxies is obtained,
it is then better to go wide. Edge effects such as loss of sensitivity
make it inefficient to consider fields which are smaller than half
a square-degree in size. Telescope start-up delays make it inefficient
to consider drift scans much shorter than a degree in length.

\subsection{{\normalsize{}Target Selection and Observations}}

The goal of AUDS is to carry out an unbiased, sensitive survey outside
the local Universe. Our original goal was to be able to detect galaxies
with $\sim0.1M_{\MHI}^{*}$  of neutral hydrogen at the frequency
limit of the ALFA receiver $(f=1225\,\mathrm{MHz})$. In order to
sample independent volumes, we chose to observe two independent fields
that contain no known clusters. Because of the long necessary integration
times, the surveyed fields are necessarily small $\sim67'\times44'$
each, corresponding to a total volume of $V\approx10^{3.9}\,\mathrm{Mpc^{3}}$.
For efficient surveying, the two regions should differ in right ascension
as much as possible, and be located at a declination where they can
be tracked for the maximum time possible at Arecibo. Another criterion
for the selection of our fields was that they were within the SDSS
survey region. We also tried to avoid bright continuum sources as
much as possible. The brightest continuum source in Field~1 is $43.9\,\mathrm{mJy}$
and in Field~2 $196.9\,\mathrm{mJy}$.

We used a ``drift and chase'' mode for the observations. Each AUDS
scan consists of 230 individual spectra for each beam and polarisation,
with each successive spectrum integrated over 1~s while the telescope
covers 1~s in right ascension. The spectra from the seven ALFA beams
were recorded using the Mock Spectrometer dividing each spectra into
two intermediate frequency (IF) sub-bands, each $172\,{\rm MHz}$
wide. The high frequency IF is centred at $f=1450\,\mbox{{\rm MHz}}$
and the low frequency IF is centred at $f=1300\,{\rm MHz}$. Combined
they cover the whole bandpass range of the ALFA receiver of $300\,{\rm MHz}$
with a spectral resolution of $\Delta f=0.02\,{\rm MHz}.$

\subsection{{\normalsize{}Data Processing }}

The bandpass removal and calibration was done using the multi-beam
single-dish data reduction software \texttt{livedata}%
\footnote{http://www.atnf.csiro.au/computing/software/livedata%
}. Details of the steps of the reduction process are described in \citet{barnes2001}.
As \texttt{livedata} was originally developed to reduce data from
the Parkes telescope for HIPASS, the program was adapted to suit the
different settings of the Arecibo telescope including the handling
of different types of FITS files and other Arecibo specific issues
such as calibration.

Each individual spectrum with 1~s integration time has a root mean
square (RMS) of about 50~$\mathrm{mJy}$. To convert the individual
spectra into regular gridded position-position-velocity cubes we use
the software \texttt{gridzilla}$^{1}$. For the final gridded data
cubes, data from all beams and polarisations were combined.

\texttt{Gridzilla} calculates the contribution of every individual
spectrum to each pixel of the grid and calculates the final value
of the pixel based on the contributing spectra and the assigned weights.
The weights were determined by the distance of spectra from the beam
centre. We used a weighted median statistic to combine the spectra.
The data values were first sorted, and their weights were summed.
The weighted median is then the data value for which the sum of the
weights is half the total weights. \citet{barnes2001} have shown
that using a median estimator was very successful for HIPASS in removing
small amounts (less then 40\%) of bad data (caused by RFI etc) with
the downside of increasing the noise level by at least $25.3\%$.

\begin{figure}
\begin{centering}
\includegraphics[width=0.5\textwidth]{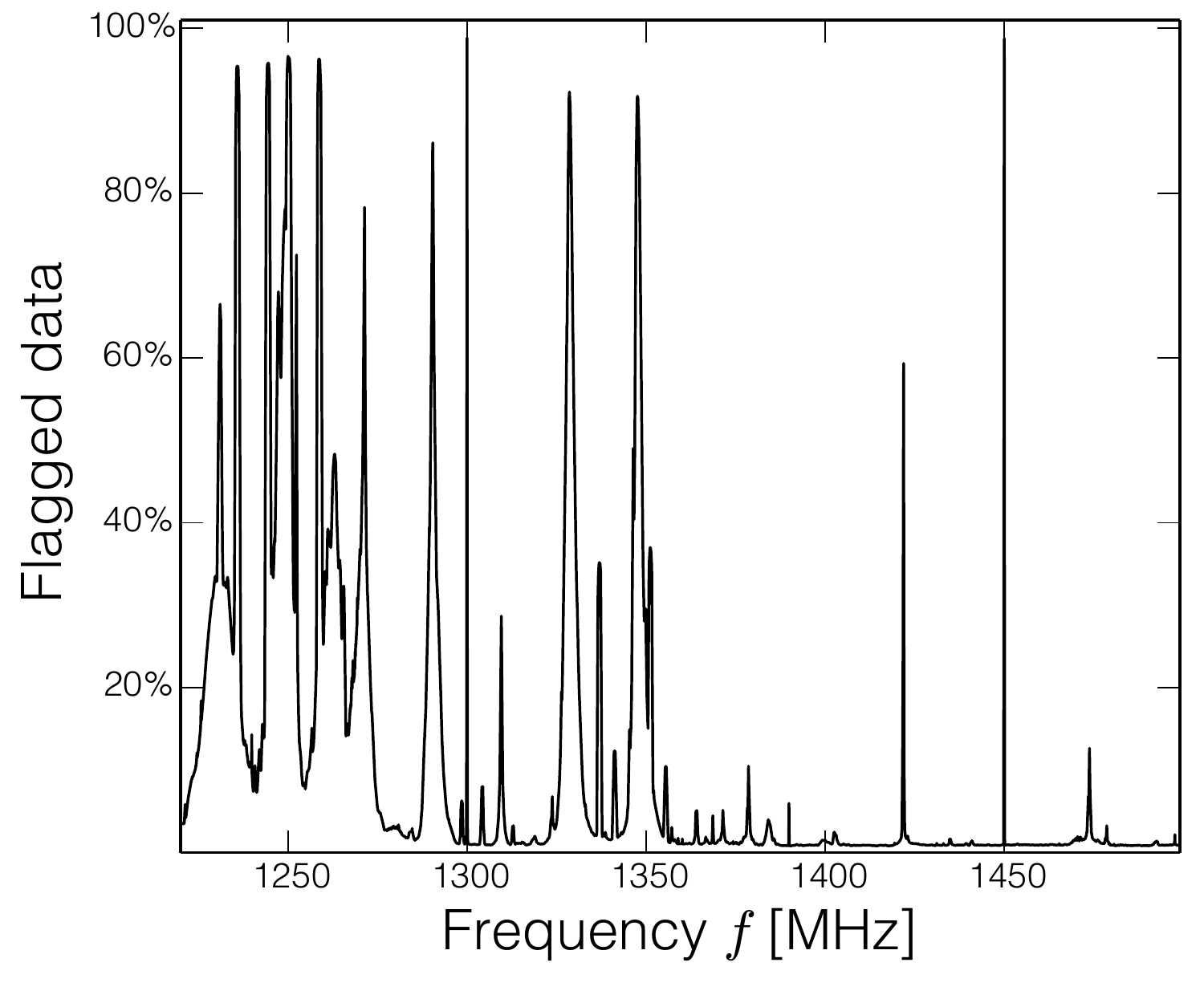}
\par\end{centering}

\protect\caption[RFI occupancy plot for AUDS.]{RFI occupancy plot for AUDS. While the higher frequency regions are
only mildly affected by RFI, large regions of the lower frequency
ranges are completely wiped out. \label{fig:RFIo_perc} }
\end{figure}

However, on occasion, the AUDS data show much more significant levels
of bad data (Figure \ref{fig:RFIo_perc}), especially in the frequency
range of $f=1220\,-\,1350\,\mathrm{MHz}$, necessitating further measures.
Several sources of strong RFI, like the radar in Punta Salinas and
transmissions from the nearby airport interfere in this frequency
range. Additionally, several satellites (e.g. GPS, Galileo and GLONASS)
transmit in this frequency range. While the RFI emitted from the radar
are narrow in frequency and pulsed in time, the RFI from the satellites
span a wide frequency range, but do not occur at all times.

\begin{figure*}
\includegraphics[width=1\textwidth]{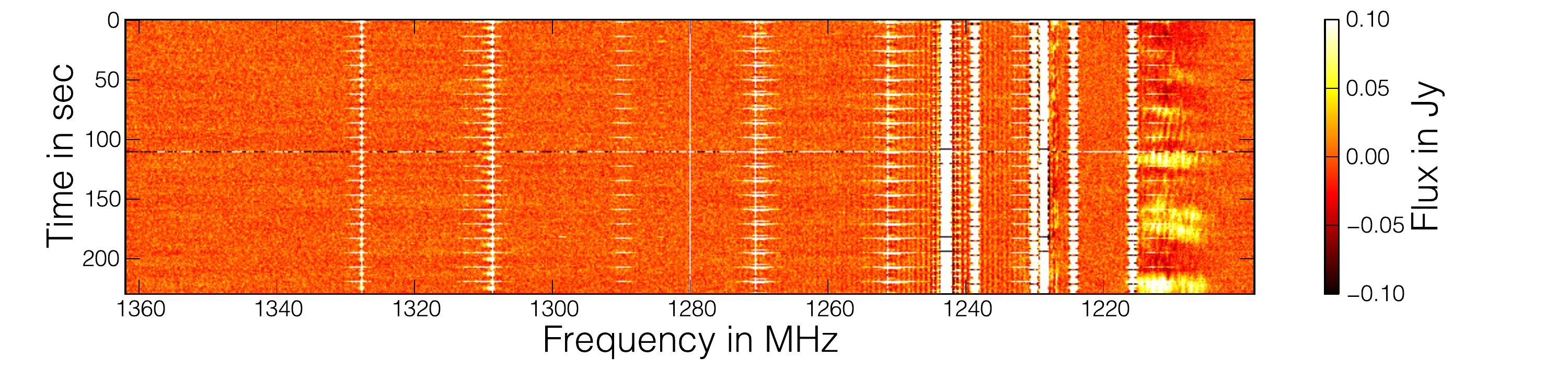} \includegraphics[width=1\textwidth]{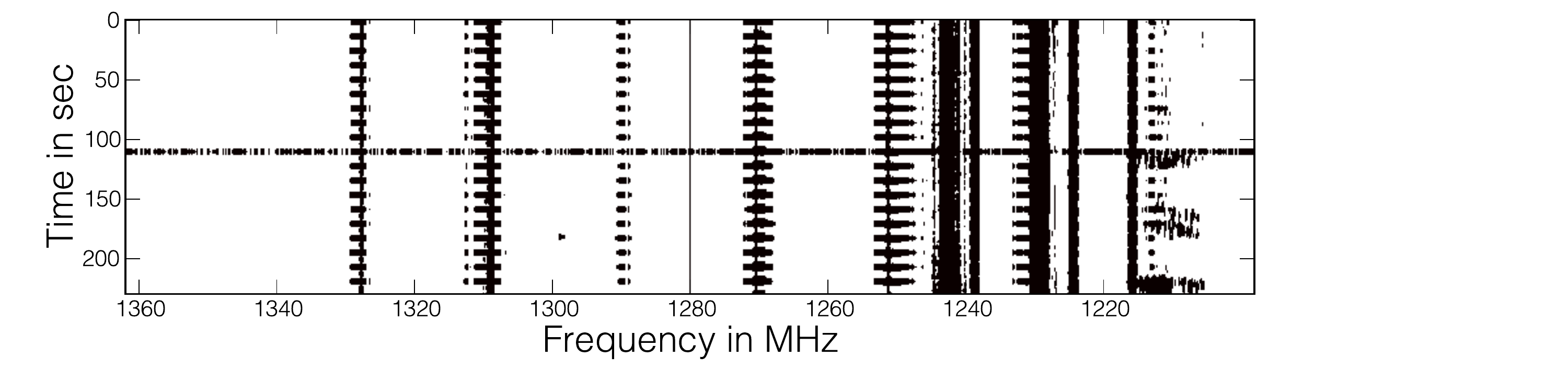}
\includegraphics[clip,width=1\textwidth]{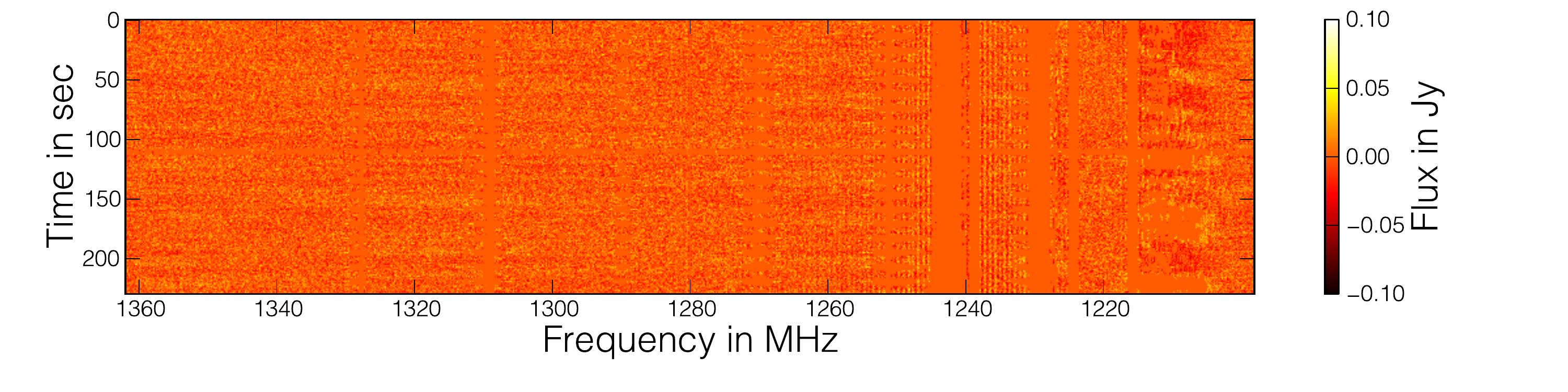}

\protect\caption[The three images show an example of the RFI flagging process.]{The three images show an example of the RFI flagging process. Upper
panel: Example of data from one beam after bandpass removal and calibration
using livedata. A stack of 230 spectra is shown with 1~s integration
time each. Centre panel: The 3$\sigma$ mask created in the flagging
process. The black regions will be masked out. Lower panel: Data after
masking with significantly reduced RFI. \label{fig:3-3Mask}}
\end{figure*}

To further mitigate against the RFI problem, we used a flagging routine
applied to the bandpass-corrected data as follows:
\begin{enumerate}
\item The RMS in a RFI-free region is estimated. The RFI free region was
chosen by examination of several time-frequency images. 
\item A $3.5\sigma$ mask is created where zero corresponds to valid pixels
and one to masked pixels. 
\item The mask is grown by smoothing it with a Hanning kernel of the size
5\texttimes 5 pixel (time-frequency domain) which results in values
between zero and one. Pixels where the values of the mask is above
0.1 are flagged and ignored during the gridding of the data. 
\end{enumerate}
This process is repeated three times before the data is gridded into
data cubes. An example of the mask and flagging is shown in Figure
\ref{fig:3-3Mask}.

\section{AUDS Sample\label{sec:3-2AUDS-Sample}}

\subsection{{\normalsize{}Galaxy Catalogue\label{sub:The-catalogue}}}

\begin{figure*}
\includegraphics[width=0.3\textwidth]{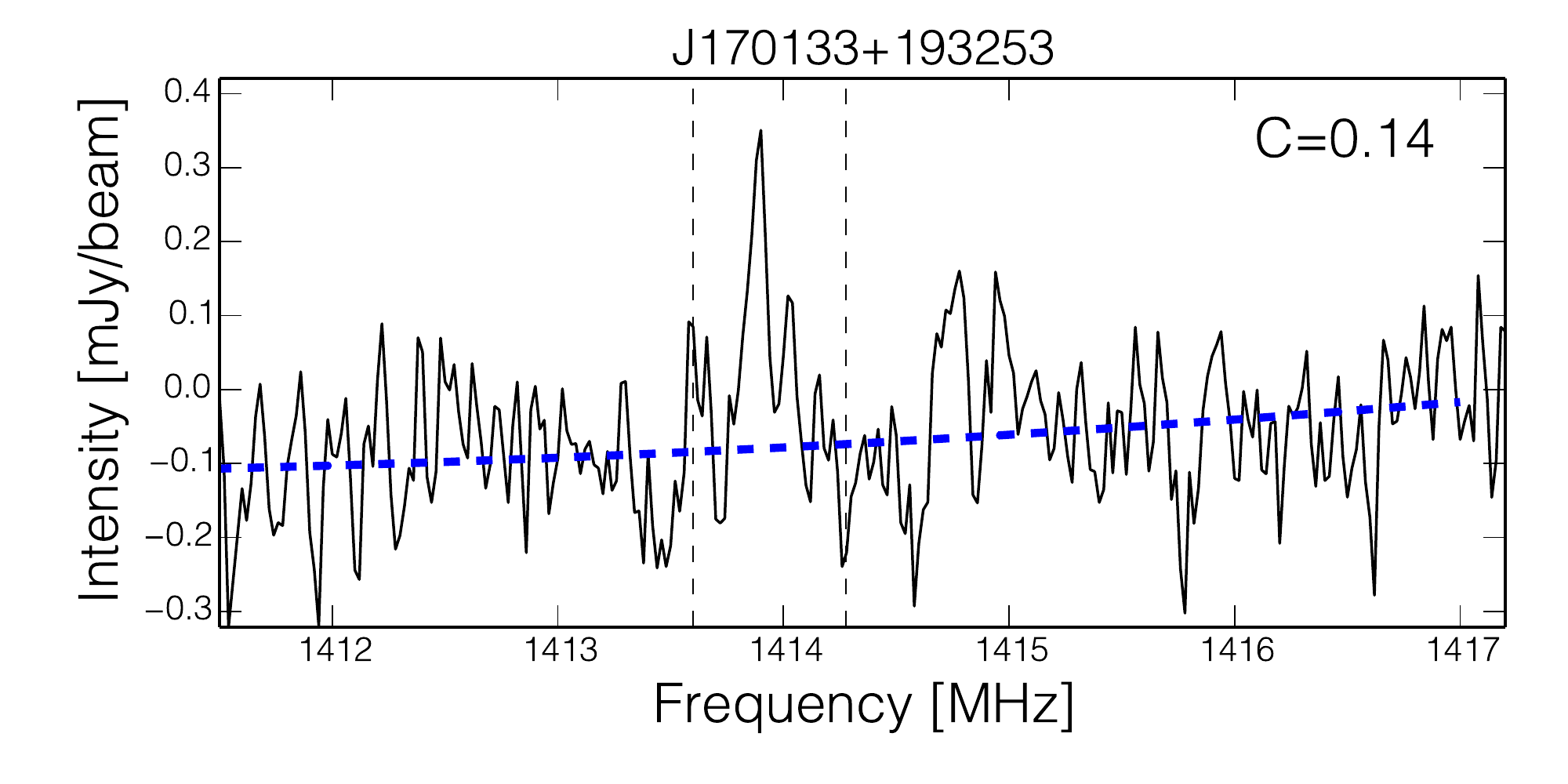}\includegraphics[width=0.3\textwidth]{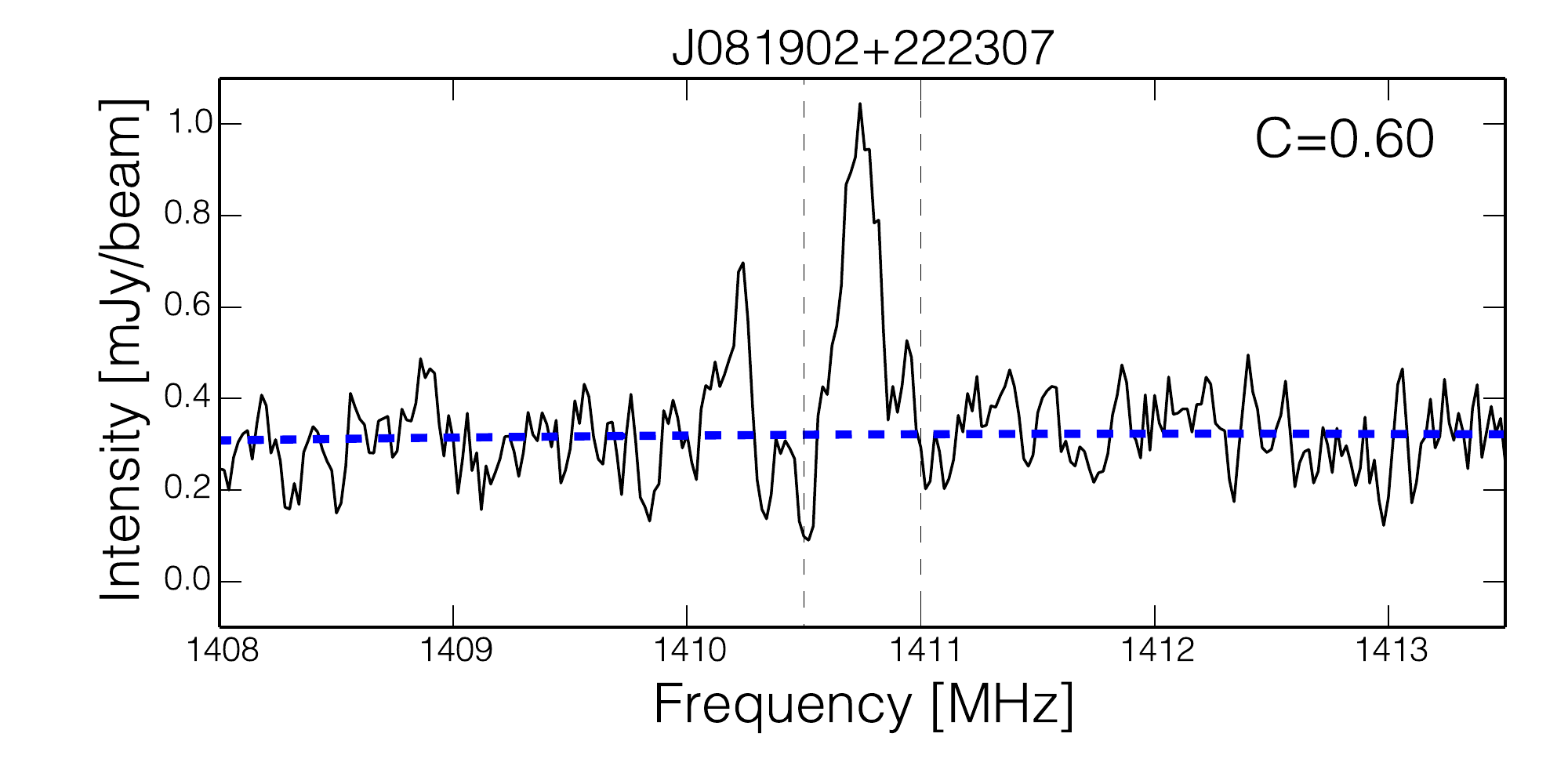}\includegraphics[width=0.3\textwidth]{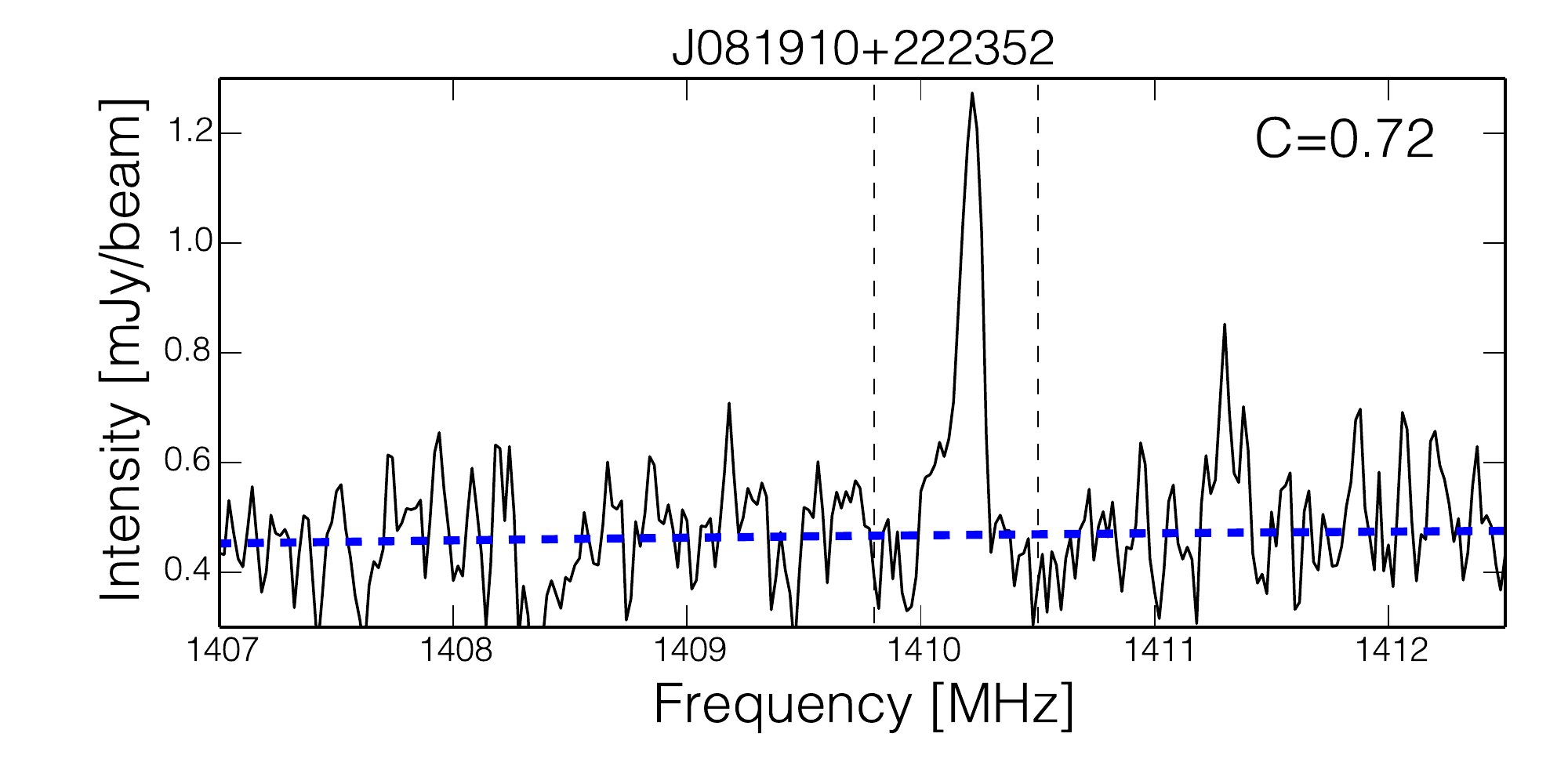}

\includegraphics[width=0.3\textwidth]{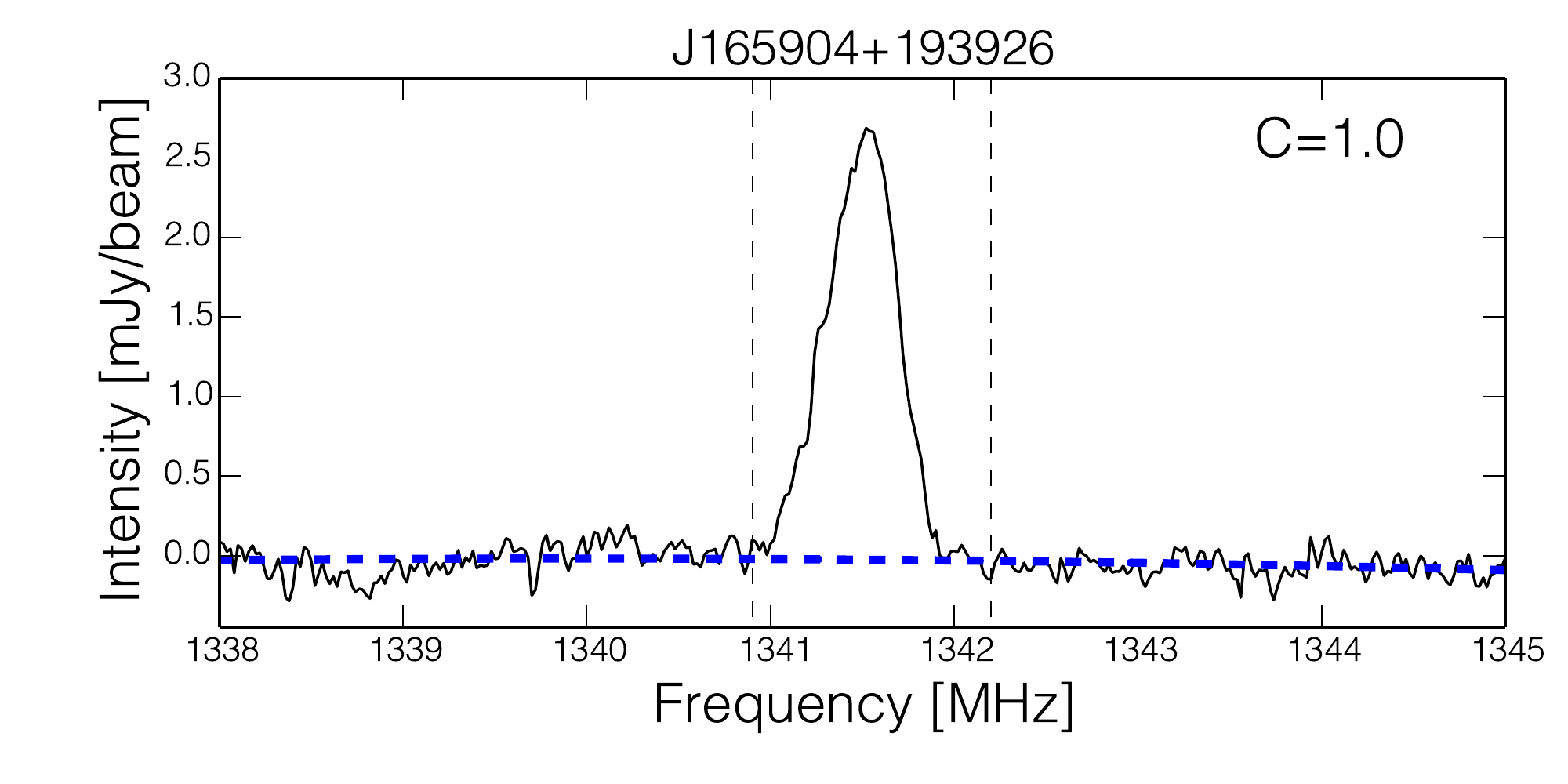}\includegraphics[width=0.3\textwidth]{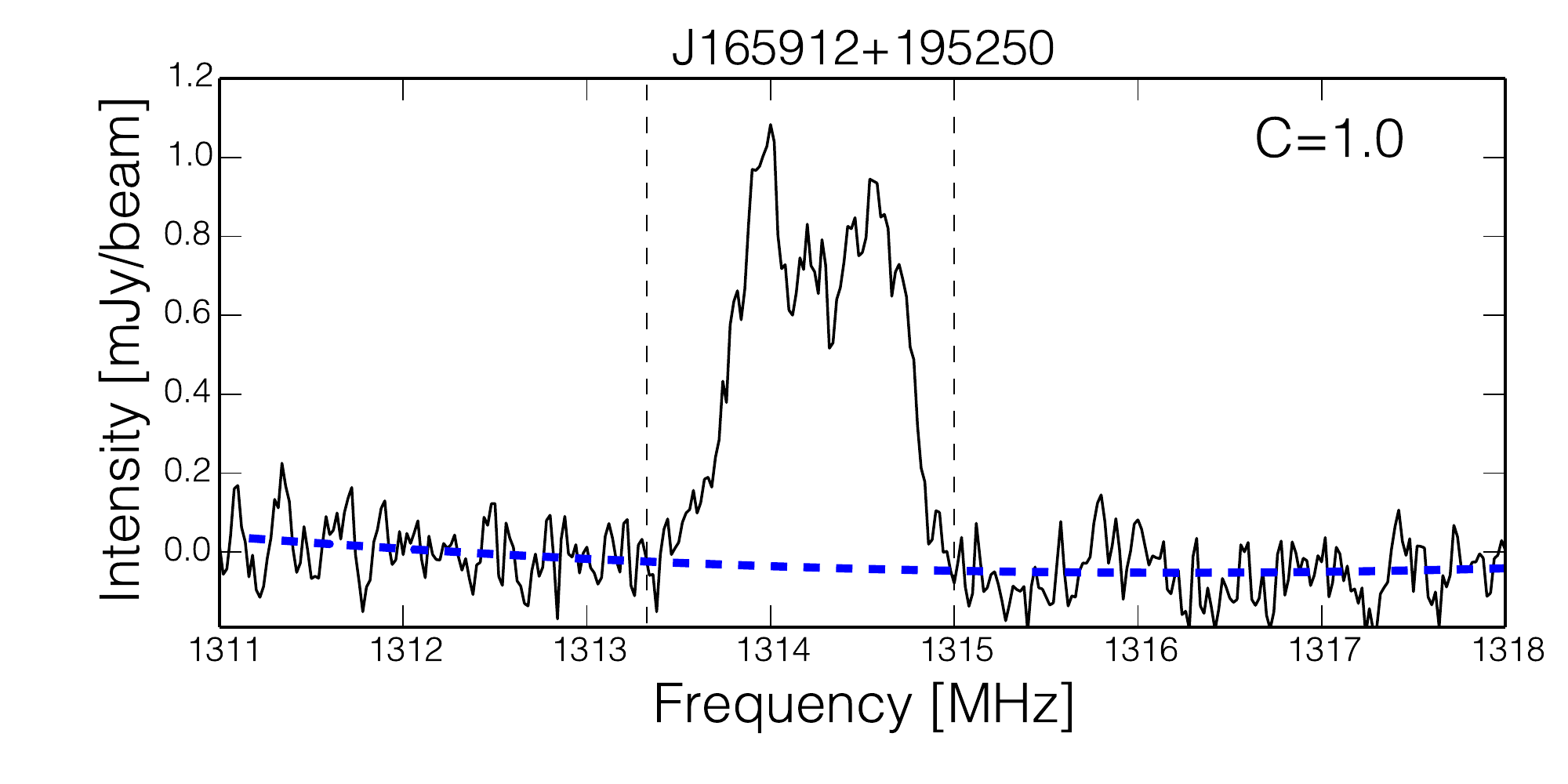}\includegraphics[width=0.3\textwidth]{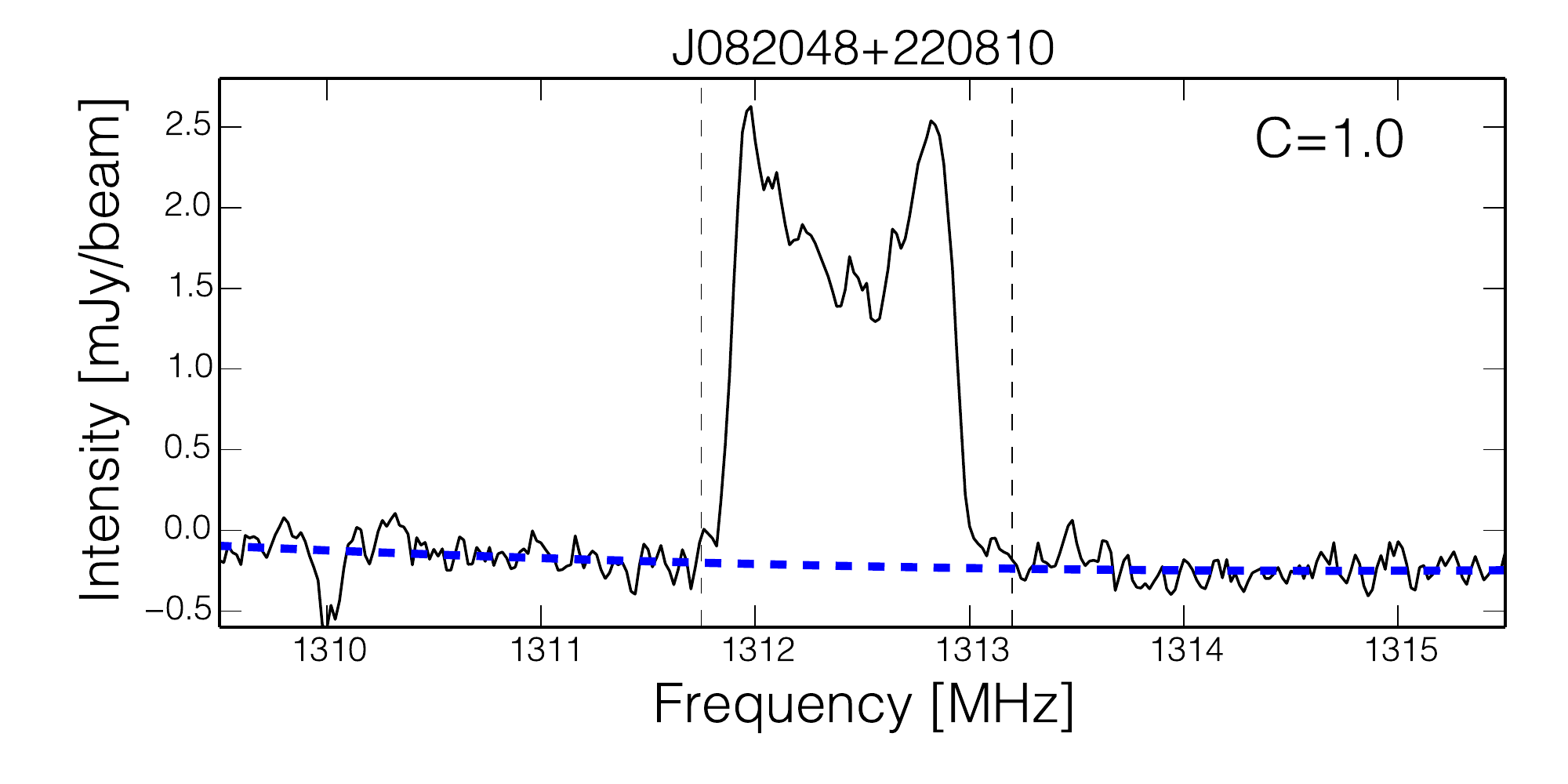}

\includegraphics[width=0.3\textwidth]{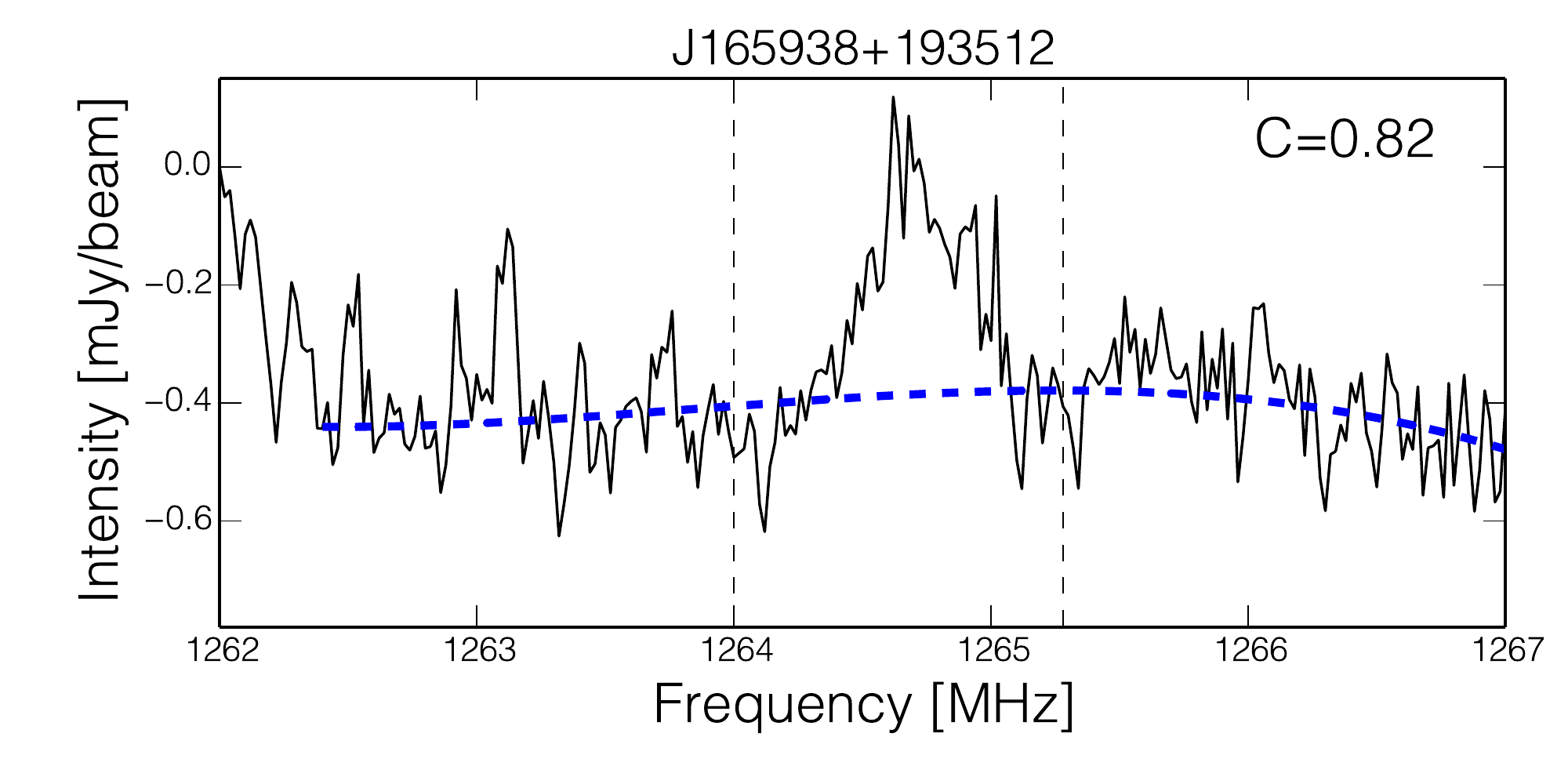}\includegraphics[width=0.3\textwidth]{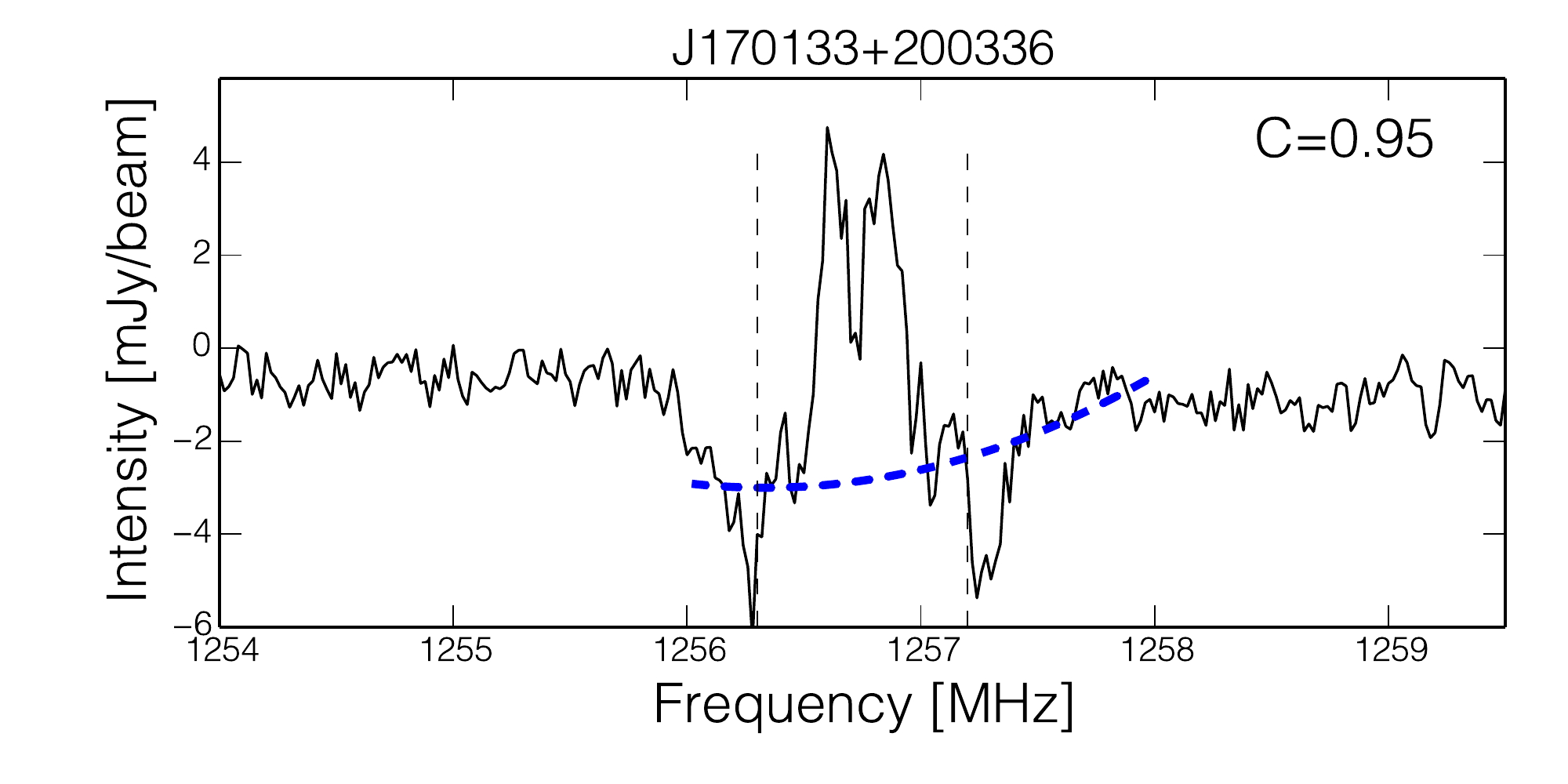}\includegraphics[width=0.3\textwidth]{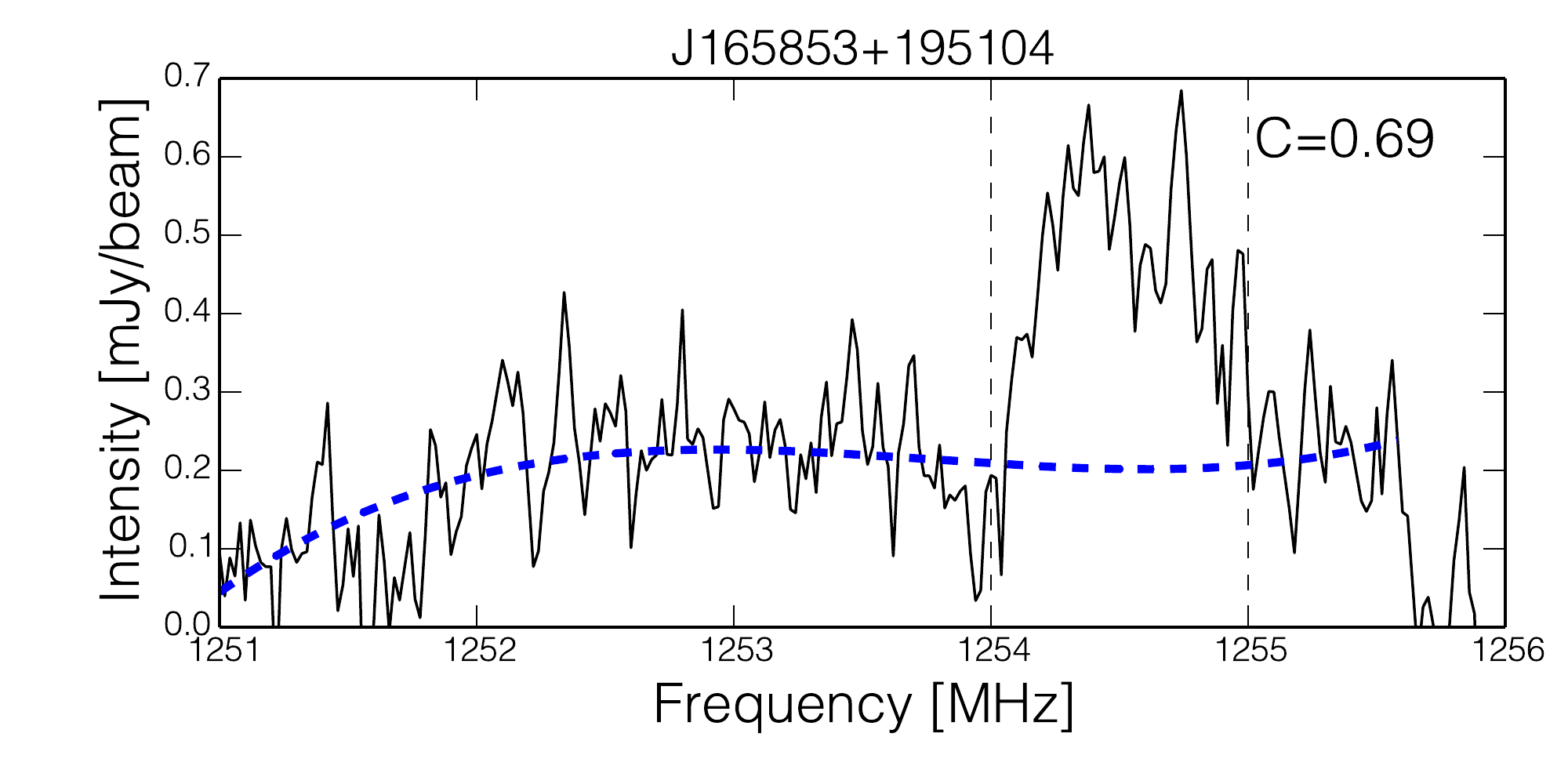}

\protect\caption[Selection of AUDS spectra. ]{Selection of AUDS spectra. The black dashed vertical lines indicate
the spectral range in which the line is fitted. This spectral region
is excluded for the fit of the baseline, marked with the blue dashed
line. The completeness coefficient for each galaxy is given in the
upper right hand corner. The top panels show the three lowest mass
galaxies detected in the survey. The central panels shows three galaxies
at intermediate redshift with high SNR. The lowest panels are the
three highest redshift galaxies. \label{fig:AUDS_spec} }
\end{figure*}

To search for galaxies, position-velocity images were searched by
eye to create a list of galaxy candidates. Due to the varying noise
and strong RFI at higher redshift, available automated source finders
were not able to create a usefully short candidate list. To create
the source list the cubes were searched by three members of the AUDS
team. Each tried to identify every possible candidate, even those
that were unlikely to be significant detections. Together, 294 unique
candidates were identified and a preliminary source list was created.
In a second step, a single person inspected each spectrum and image
in position\textendash velocity space from the merged list. Spectra
were extracted and fitted and a shorter source list created based
on the final inspection of each candidate. It was important to carefully
distinguish between real galaxies and spurious detections based on
their signal-to-noise ratio (SNR), their spectral line shape and the
shape of the detections in the image plane. Special attention was
taken to distinguish between real sources and RFI. In many cases a
clear distinction was possible given their very different signatures
in the position-velocity plane, RFI often being narrow in frequency
and visible over large ranges in right ascension/declination. However,
at low SNR, the distinction is not as clear. Candidates close to,
or overlapping with known RFI were treated especially carefully to
ensure they were real and that the bandpass was fitted correctly.
This resulted in a shorter list of 133 likely candidates, five of
which were common between the overlapping bands leaving 128 candidates.

We then used the completeness coefficient $C$, derived from the simulation
discussed in Section \ref{sub:Completeness}, to create a final source
list. $C$ uses the integrated flux $\Sint$, the velocity width $W$
and the RMS $\sigma$, of each galaxy to calculate the probability
of detecting a galaxy in the survey. If $C>0$ the galaxy was retained
in the sample.

The final sample consists of\foreignlanguage{english}{ $\NNo$ galaxies.
A selection of spectra of AUDS galaxies is presented in Figure \ref{fig:AUDS_spec}.}
Parameterisation of the galaxies was performed using the task \texttt{mbspect}
which is part of the radio astronomy data reduction package\texttt{
miriad} \citep{Sault1995ASPC...77..433S}%
\footnote{http://www.atnf.csiro.au/computing/software/miriad/ %
}. The galaxy position estimated from the manual search was more precisely
determined by fitting a Gaussian to the velocity integrated image
(0th moment) over a width of 5-9 arcmin around the input position
assuming the galaxies are point sources. This is a good approximation
as the beam size at the mean redshift of the galaxies in the sample
($z=0.065)$ is about $190\, h^{-1}\,{\rm kpc}$ in diameter. Additionally
we also looked at the optical diameter (Petrosian diameter in the
$r$-band) of the AUDS galaxies we could cross match to SDSS galaxies
(details are given in Section \ref{sub:Optical-Counterpart}) and
found that 95\% of galaxies are smaller than 1.2~arcmin. This would
allow the $\hi$ disk to exceed the optical diameter by up to three
times and still be within the ALFA beam. 

\begin{figure}
\centering{}\emph{\includegraphics[width=1\columnwidth]{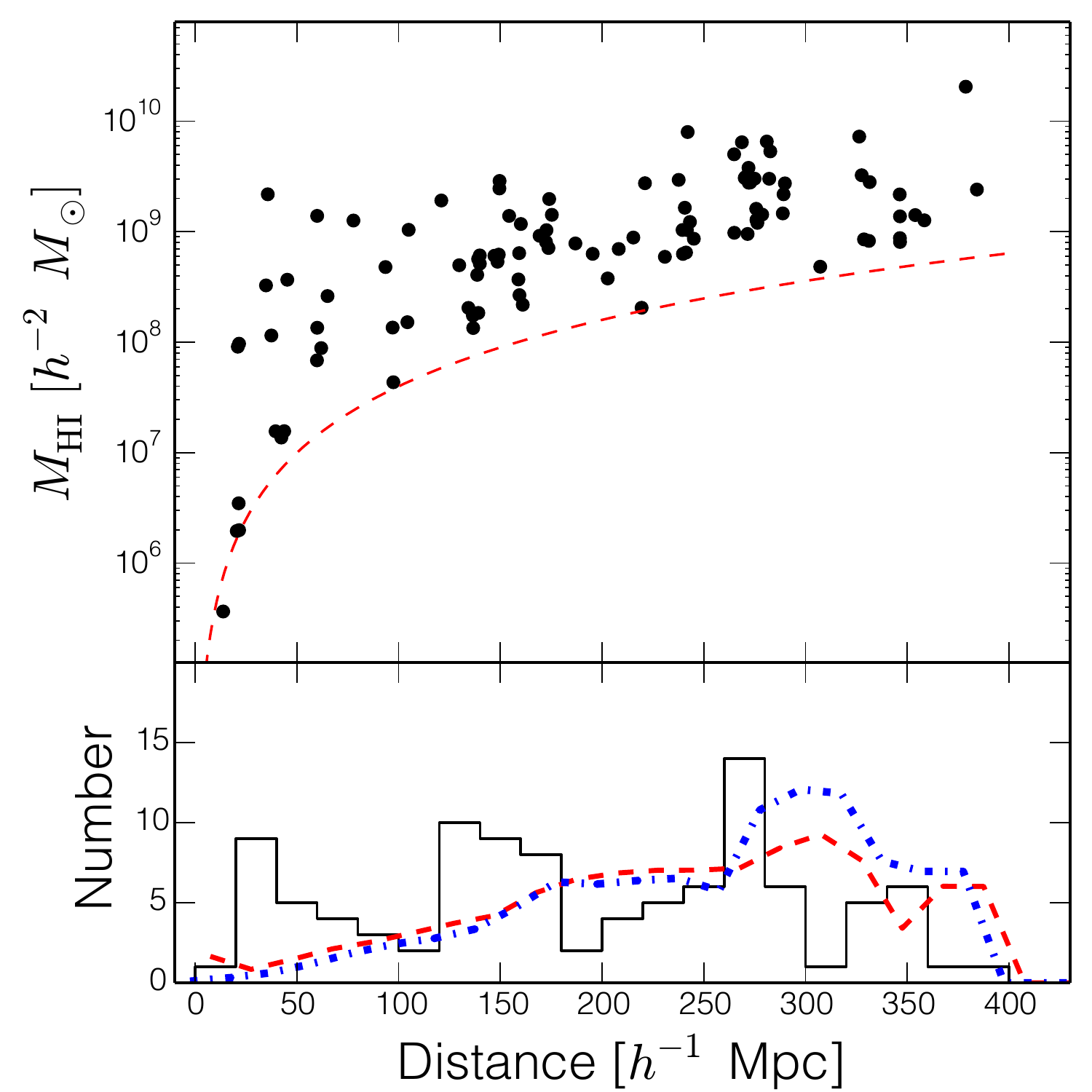}\protect\caption[$\hi$ masses of the AUDS galaxies as a function of distance (top
panel) and the distribution of detected galaxies in comparison to
the 2DSWML and $\sum\,1/\Vmax$ selection function. ]{Top panel: $\hi$ masses of the AUDS galaxies as a function of distance.
The red dashed line indicates the minimal detectable $\hi$ mass as
a function of distance, assuming an integrated flux limit of $\Sint=0.8\,\mathrm{mJy\, MHz}$.
Bottom panel: The black, solid histogram shows the distribution of
detected galaxies. The red dashed line is the expected galaxy distribution
derived by multiplying the 2DSWML selection function by $\Omega D^{2}\Delta D\overline{n}$.
The blue dash-dotted line is the prediction using the selection function
from the $\sum\,1/\Vmax$ method. \label{fig:MHI_hist} }
}
\end{figure}

Next the spectra were optimally extracted, also using the \texttt{miriad}
function \texttt{mbspect}, at the new position using a window of 5-9
arcmin, weighting neighbouring pixels by the beam shape. The velocity
range occupied by the detected galaxy was masked out, before a polynomial
was fitted to the baseline. The spectral width $W$ and the central
velocity of the profiles were measured. The fit also provided $\Sint$,
$S_{\mathrm{peak}}$ and the peak SNR. The fluxes were measured assuming
the galaxies are point sources. Using $\Sint$ and the luminosity
distance $D_{\mathrm{L}}$ of the galaxies we calculate $M_{\hi}$

\begin{equation}
\frac{{M_{\mathrm{{\hi}}}}}{M_{\odot}}=49.8\;\bigg(\frac{{D_{\mathrm{{L}}}}}{\mathrm{Mpc}}\bigg)^{2}\;\bigg(\frac{{S_{\mathrm{{int}}}}}{\mathrm{Jy\, Hz}}\bigg).
\end{equation}
The $\hi$ masses of the AUDS galaxies as a function of their distance
is presented in Figure \ref{fig:MHI_hist}. 
 
\begin{figure*}
\includegraphics[width=0.9\textwidth]{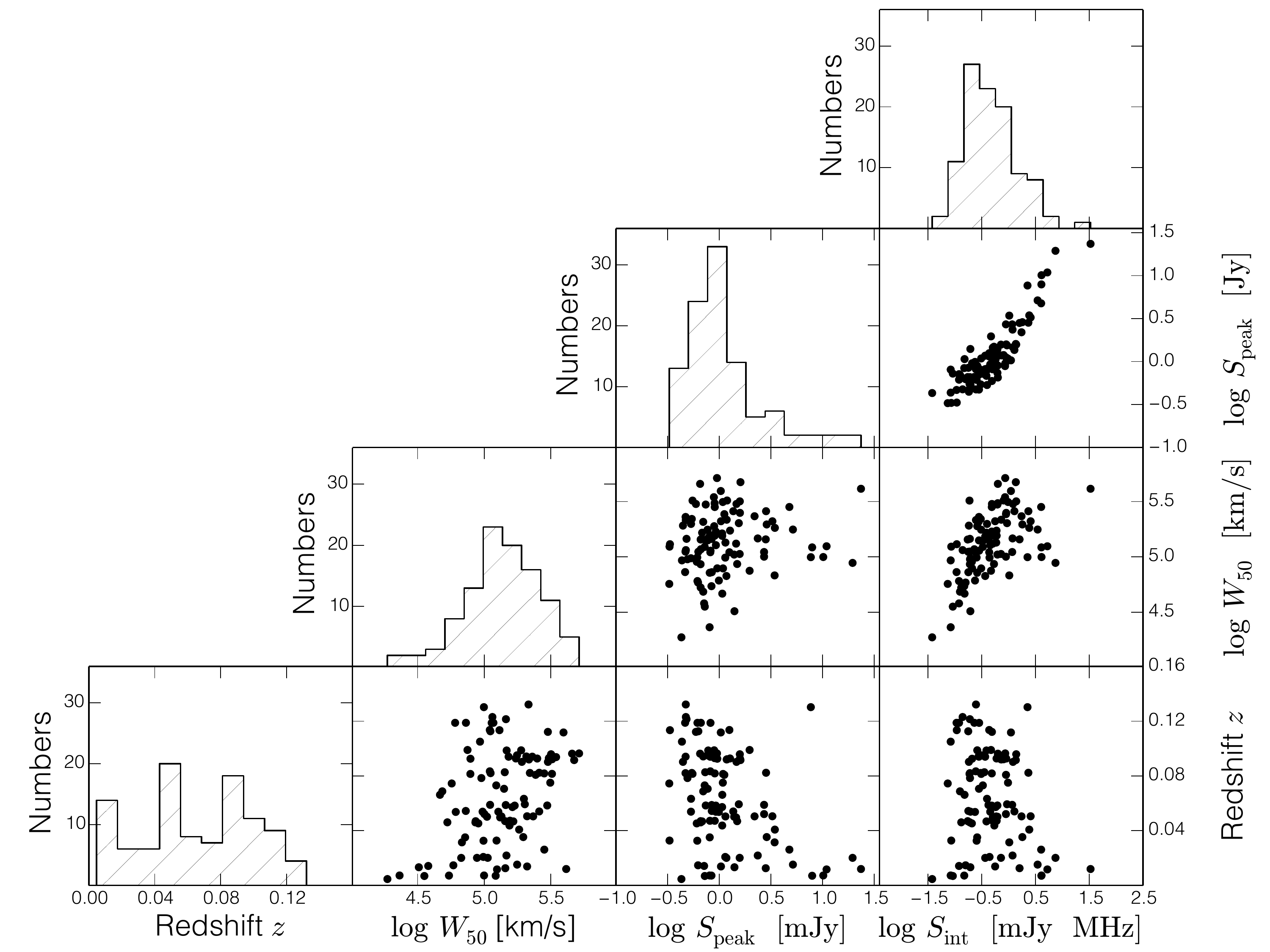}

\protect\caption[Log-log bivariant distribution of measured parameters of AUDS galaxies. ]{Log-log bivariate distribution of measured parameters of AUDS galaxies:
redshift $z$, 50\% velocity width $W_{50}$, peak flux $S_{\mathrm{peak}}$
and integrated flux $\Sint$. On the diagonal the single-parameter
histograms are plotted.\label{fig:par}}
\end{figure*}
 The bivariant distributions as well as single parameter histograms
of the galaxy parameters ($z$, $\Sint$, $S_{\mathrm{peak}}$, $W$)
are shown in Figure \ref{fig:par}. The redshift histogram shows that
we did not detect any galaxies with $z>0.13$. Detecting $\hi$ galaxies
at high redshift proved challenging as the RFI environment at Arecibo
is very hostile, especially for $f<1290\,\mathrm{MHz}$.

\subsection{{\normalsize{}Optical Counterparts \label{sub:Optical-Counterpart}}}

Both AUDS fields were chosen to overlap with the SDSS footprint. Searching
the SDSS DR7 catalogue we find Field 1 has 25 galaxies and Field 2
has 56 galaxies with spectroscopic redshifts in the redshift range
of AUDS. Additionally there are 11588 galaxies with only photometric
redshifts in Field 1 and 9932 galaxies in Field 2.

\begin{figure*}
\includegraphics[width=0.45\textwidth]{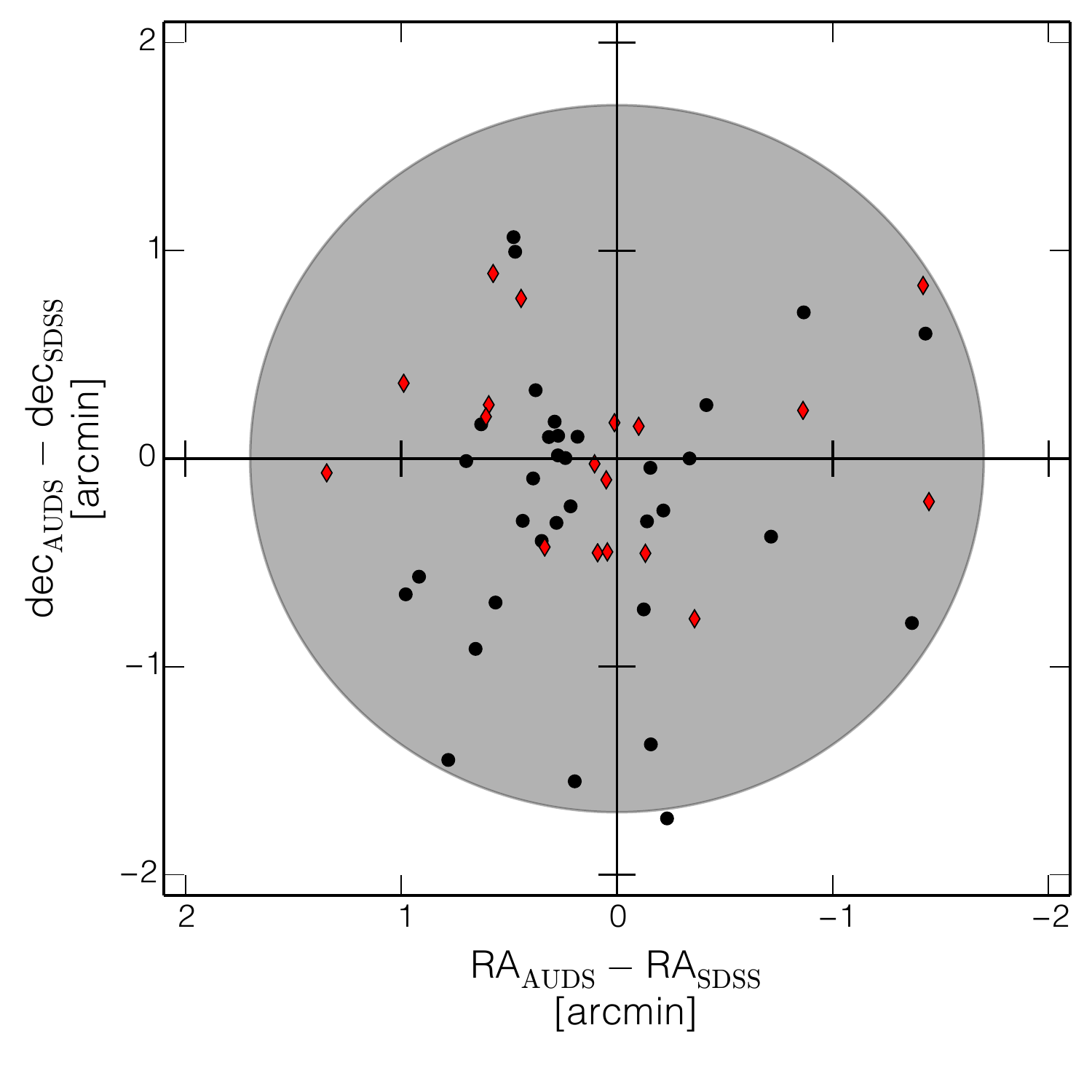} \includegraphics[width=0.45\textwidth]{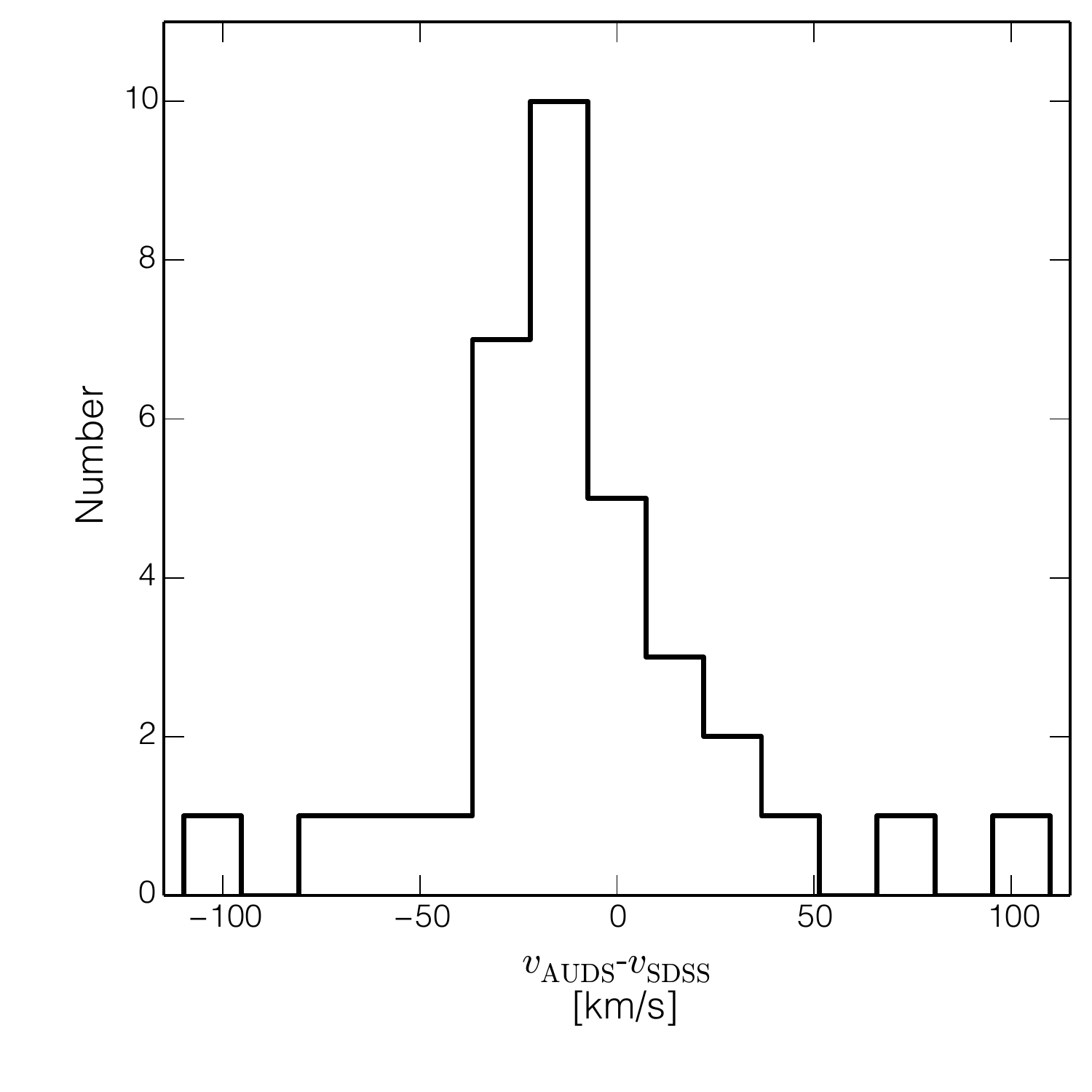}\protect\caption[Position of AUDS galaxies with spectroscopic counterparts and Histogram
of the difference in redshift between the measured optical velocity
from SDSS and the AUDS. ]{\label{fig:RAerr}Left panel: Black dots: AUDS galaxies with spectroscopic
counterparts, Red diamonds: likely optical counterparts with only
photometric redshifts. The plot shows the difference between the source
position fitted to the AUDS galaxies and the source position given
in the SDSS catalogue. The grey shaded region is the size of the ALFA
beam. Right panel: Histogram of the difference in redshift between
the measured optical velocity from SDSS and the AUDS for all galaxies
with spectroscopic counterparts. }
\end{figure*}

We used the SDSS to find optical counterparts for the AUDS galaxies
and found that 36 out of $\NNo$ have spectroscopically confirmed
counterparts and that at least one of the AUDS galaxies is a pair.
Optical galaxies are identified as matches when they are within a
two arcmin radius around the $\hi$ position and the difference with
the spectroscopic redshift is smaller than $150\,\mathrm{km\, s^{-1}}$.
Taking the position (inside the beam), the size, inclination and colour
of the galaxy into account we identify another 18 likely optical counterparts.
Figure \ref{fig:RAerr} (left panel) shows the offset between the
position of the optical counterparts as given by the SDSS and the
positions measured from the $\hi$ data for both the galaxies with
optical counterparts with spectroscopic redshifts as well as the galaxies
with likely counterparts with photometric redshifts. We also show
the difference between the optical redshift and $\hi$ redshift in
the right panel of the same figure. 

\begin{figure*}
\begin{centering}
\includegraphics[bb=25bp 0bp 425bp 260bp,clip,height=5cm]{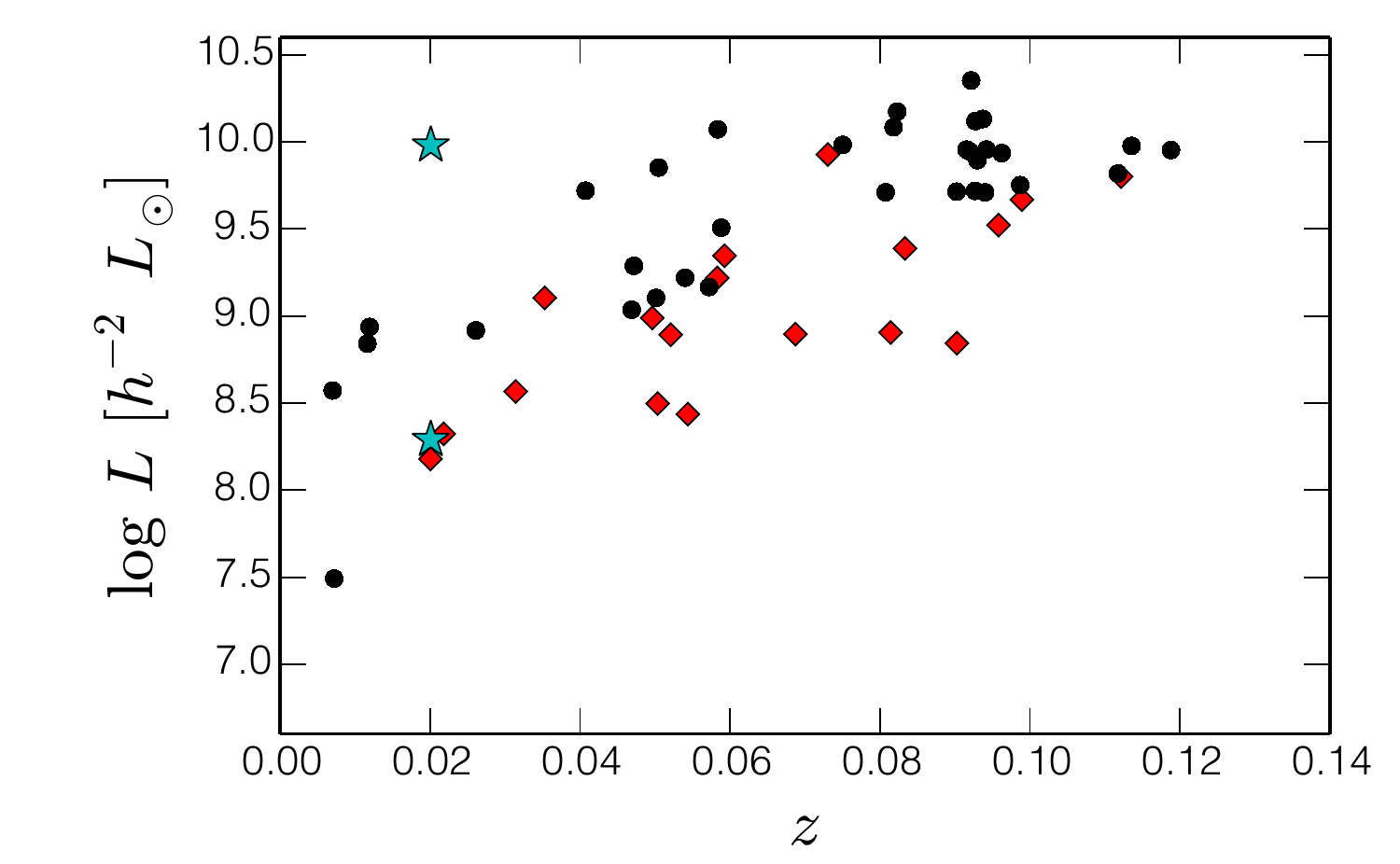}\includegraphics[bb=25bp 0bp 425bp 270bp,clip,height=5cm]{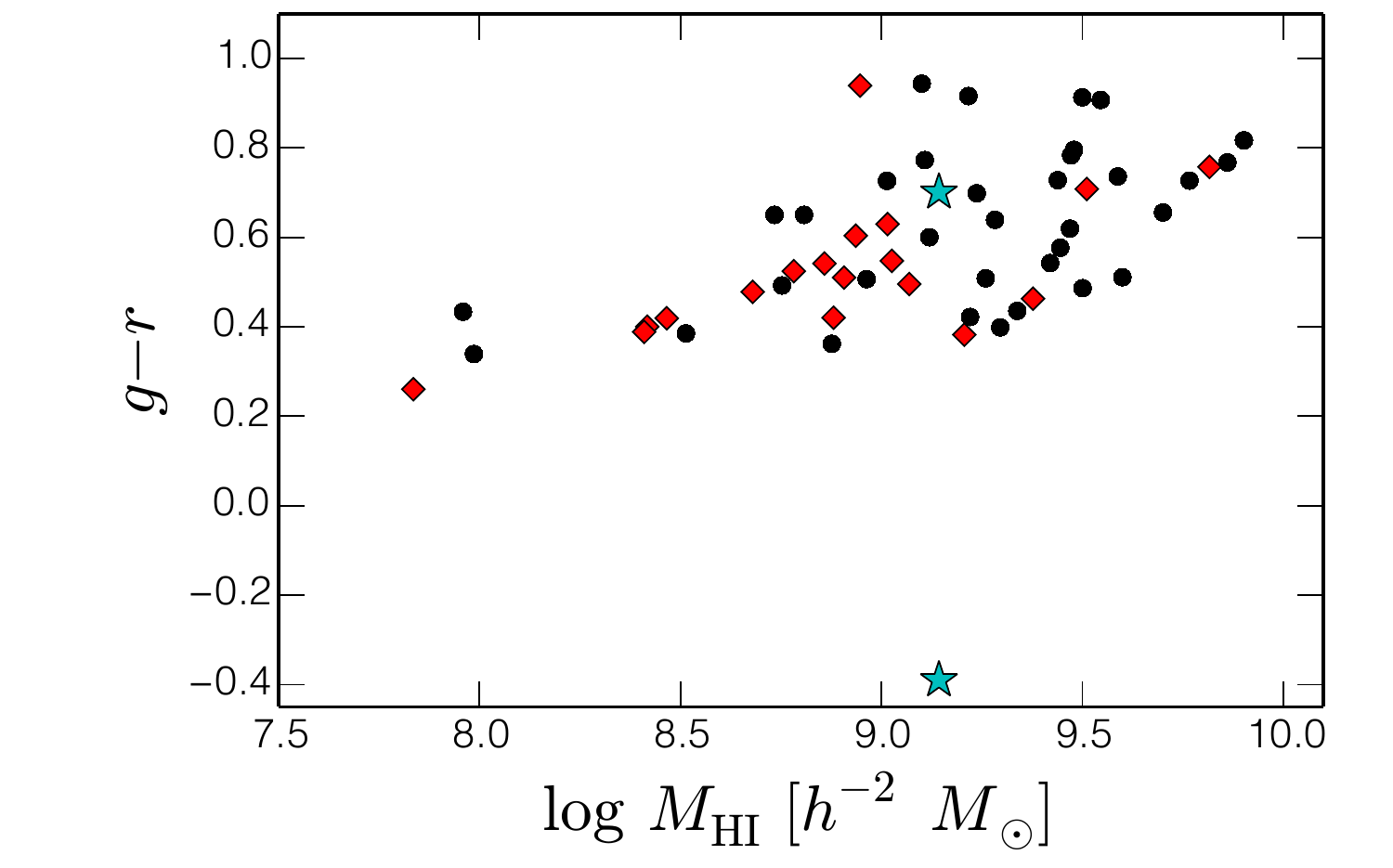}
\par\end{centering}

\protect\caption[Optical counterparts of AUDS galaxies.]{\label{fig:Opt_Count} Optical counterparts of AUDS galaxies. Black
dots: $\hi$-selected AUDS galaxies with optical spectroscopic counterparts;
red diamonds: likely candidates with only photometric redshift measurements;
cyan stars: galaxies in a pair. Left plot: AUDS is not volume limited
and therefore we find a correlation between the redshift and the detected
luminosity of the galaxy. Less luminous galaxies (at a certain redshift)
are less likely to have spectroscopic counterparts. Right plot: Lower
mass galaxies are predominantly blue galaxies $(g-r<0.7)$. Redder
galaxies are only found at larger $\MHI$. As bluer galaxies tend
to be more gas rich they are more easily picked up at low masses in
comparison to red galaxies which have lower gas mass fractions but
are significantly more massive and luminous and therefore easier detectable
at higher redshifts by optical surveys. }
\end{figure*}

Even though the overall number of galaxies with reliable optical information
is small, some trends are clearly visible (Figure \ref{fig:Opt_Count}).
There is a correlation between the detectability of a galaxy and its
$r$-band luminosity as the SDSS is a magnitude but not volume-limited
sample. This also becomes evident when one compares the luminosity
of the optical counterparts with and without spectroscopic redshifts.
Galaxies with higher masses at a certain redshift are more likely
to have spectroscopic counterparts.

The second trend is that lower mass galaxies are predominately blue
galaxies $(g-r<0.7)$ and less likely to have spectroscopic information.
Redder galaxies are only found at large $\MHI$. This is expected
as bluer galaxies tend to be more gas rich. Red galaxies on the other
hand tend to have lower gas mass fractions but are significantly more
massive and luminous. That makes them easier to detect in optical
surveys especially at higher redshift where the survey volume is larger.
This means that non-volume limited $\hi$ surveys tend to be biased
against red galaxies with increasing redshift. We leave the more detailed
discussion of optical properties as well as a stacking analysis for a
later paper. 

\subsection{{\normalsize{}Survey Completeness\label{sub:Completeness}} }

The completeness of a survey is defined as the fraction of a certain
type of galaxy which can be detected by a survey from the underlying
distribution of objects down to the detection limit of the survey.
Estimating the completeness of AUDS is a crucial step to understand
the underlying galaxy population. 

\begin{figure*}
\includegraphics[width=0.5\textwidth]{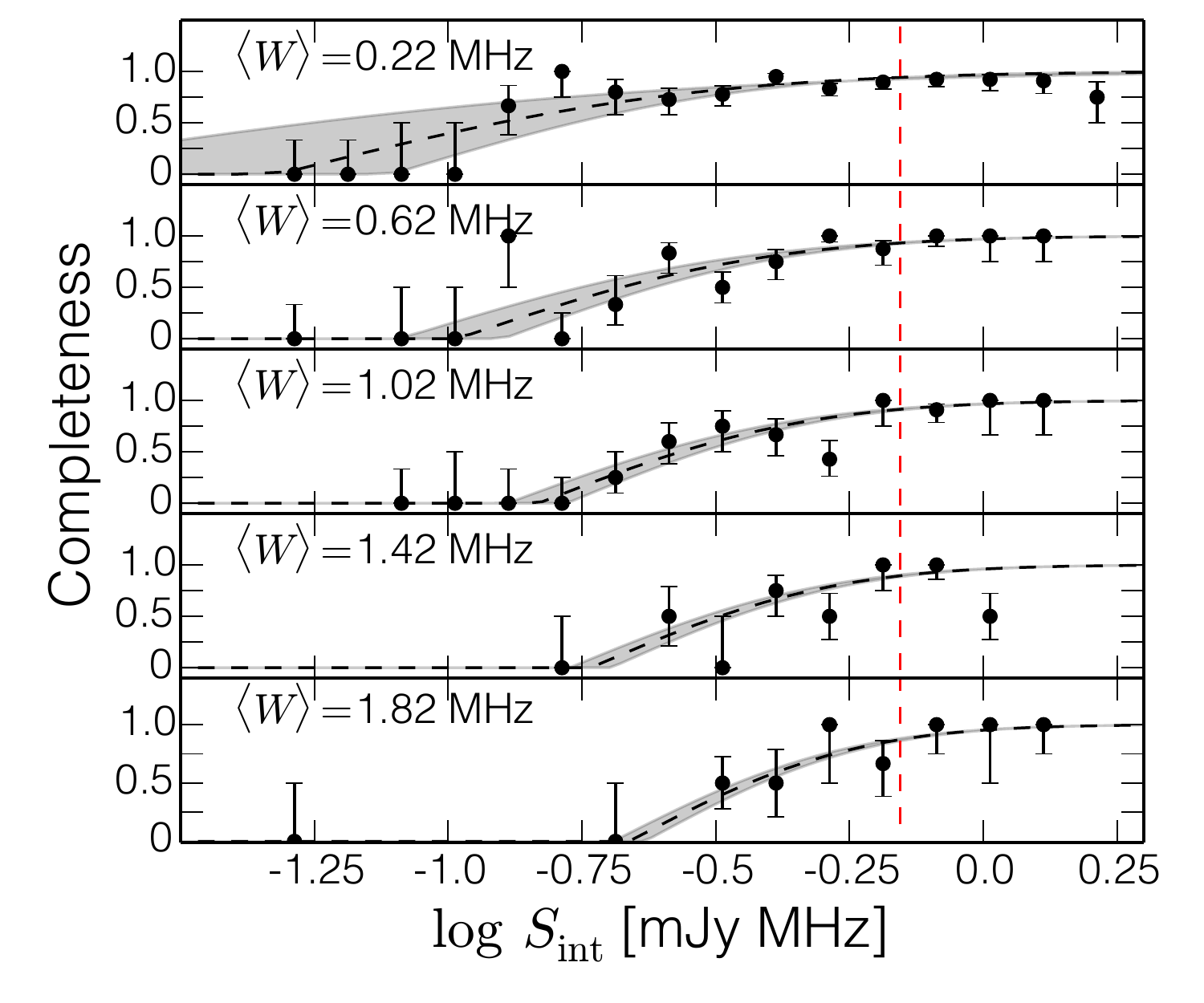}\includegraphics[width=0.5\textwidth]{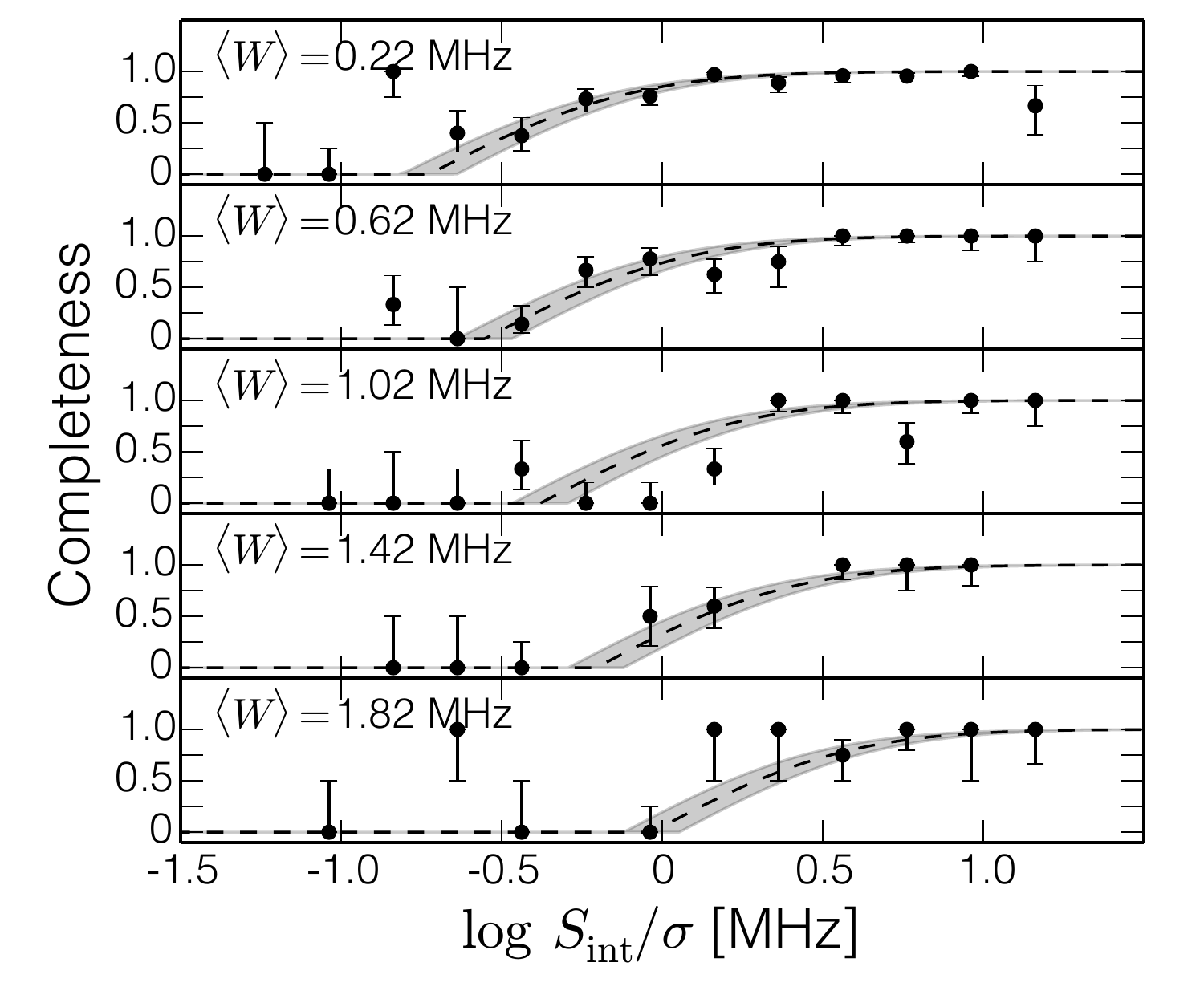}

\protect\caption[1d-slices of the completeness function in different bins of frequency
width.]{\label{fig:Comp3d} 1D-slices of the completeness function in different
bins of frequency width. The central width of each bin is given in
the upper left hand corner. The completeness is well described by
an error function shifted along the horizontal axis. The grey shaded
area indicates the completeness for the maximum and minimum width
in the bin. Left panel: The completeness as a function of $\Sint$
and $W$. The dashed line is the best-fit to the data using equation
\ref{eq:Cross2-1}. The red dashed line indicates a nominal completeness
limit of $\Sint=0.8\,\mJyMHz$.  Right panel: The completeness as
a function of $\Sints$ and $W$. The dashed line is the best-fit
to the data using equation \ref{eq:Cross2-1-1}. }
\end{figure*}

A good way to determine the completeness of the sample is to insert
synthetic galaxies into the data and test the rate with which these
galaxies are detected. This allows us to determine a completeness
limit as a function of $\Sint$ as well as $W$ and the single pixel
noise $\sigma$. To take the noise into account we defined a noise
weighted integrated flux $\Sints=\Sint/\sigma$.

To insert the galaxies we chose three representative subregions in
the data cube of Field 1: one RFI free region ($1381\,\MHz$); one
region moderately affected by RFI $(1325\,\mathrm{MHz)}$; and one
region with high RFI occupancy $(1231\,\mathrm{MHz)}$. Each region
is $20\,\mathrm{MHz}$ wide. In total 356 synthetic galaxies were
inserted into the subcubes. The parameters for the synthetic galaxies
were randomly chosen from an integrated flux range of $S_{\mathrm{{int}}}=0$
to $1.5\,\mathrm{{mJy\, MHz}}$ and a width range $ $of $W=0.12$
to $2.0\,\mathrm{{MHz}}$. Varying the width of the galaxies takes
both variation due to different rotational velocities as well as different
inclination into account. The shapes of the galaxy spectra were described
by a busy function \citep{Westmeier2014}.

The data cubes were searched blindly in the same way as the original
data cubes and 256 of the galaxies in the mock sample could be recovered.
From the ratio of the number of detected galaxies $(n_{\mathrm{d}})$
to the number of all galaxies $(n_{\mathrm{t}})$ the completeness
in logarithmic flux and linear $W$ bins can be computed.

The dependency of the completeness on $W$ and $\Sint$ as well as
$W$ and $\Sints$ ,shown in Figure \ref{fig:Comp3d} resembles an
error function shifted along the flux axis. We chose to describe the
completeness function with the following analytic model

\begin{equation}
C=\mathrm{max}\Bigg(\mathrm{erf}\bigg(\alpha W^{\alpha}(\Sint-\beta W-\gamma)\bigg)\,,\,0\Bigg)\label{eq:Cross2-1}
\end{equation}

\begin{equation}
C=\mathrm{max}\Bigg(\mathrm{erf}\bigg(\alpha W^{\alpha}(\Sint^{*}-\beta W-\gamma)\bigg)\,,\,0\Bigg).\label{eq:Cross2-1-1}
\end{equation}
The four independent parameters in equation \ref{eq:Cross2-1} and
\ref{eq:Cross2-1-1} were chosen to fit the features in the binned
completeness data namely: (1) the completeness decreases for smaller
fluxes for all values of $W$; (2) $W$ influences the steepness of
the decrease of the completeness to zero (narrow profiles are easier
to detect) and (3) the flux value at which the completeness essentially
becomes zero.

The errorbars in Figure \ref{fig:Comp3d} indicate the $1\,\sigma$
binomial confidence interval given by $n_{\mathrm{d}}/n_{\mathrm{t}}\,\pm\,\sqrt{(p(1-p)/n_{\mathrm{t}})}=p$
with $ $$n_{\mathrm{d}}$ being the number of detected synthetic
galaxies in a bin, $n_{\mathrm{t}}$ the number of all synthetic galaxies
in that bin and $p$ the upper/lower limit of the confidence interval.
Binning data in intervals of $\Delta S_{\mathrm{int}}=0.1\,\mJyMHz$
and $\Delta W=0.4\,\mathrm{MHz}$ and fitting equation \ref{eq:Cross2-1}
gives the best-fit parameter of $\alpha=28.95,$ $\beta=-0.08$, $\gamma=-5.21$
and $\delta=0.29$. Using the noise weighted flux ($\Delta\Sints=0.2$,
$\Delta W=0.4\,\mathrm{MHz}$) and fitting equation \ref{eq:Cross2-1-1}
gives the best-fit parameter of $\alpha=68.39,$ $\beta=0.45$, $\gamma=-0.52$
and $\delta=0.030$. Since the noise changes throughout the cube,
we found that equation \ref{eq:Cross2-1-1} is the better way to minimise
the effect of the different noise levels and we therefore used $C(\Sints,W)$
when calculating the HIMF. For each galaxy in the sample we calculate
the completeness coefficient by inserting the value for $\Sint$,
$\sigma$ and $W$ into equation \ref{eq:Cross2-1-1} giving us a
specific value for that galaxy which represents the probability of
detecting this galaxy in the sample. Additionally we give the integrated
flux at which a galaxy of a certain width reaches a completeness of
$C=95\%,\,90\%,\,65\%$ in Table \ref{tab:3-2-WSInt}. As before,
only galaxies with completeness coefficients $ $$>0$ have been accepted
in the sample (see Section \ref{sub:The-catalogue}). Some of the
galaxies with a completeness coefficient $ $$=0$ may in fact be
real, but it is not possible to use them for statistical studies.
\begin{table}
\centering{}\protect\caption{This table summarises the integrated fluxes at which a galaxy with
a certain width reaches a completeness limit of 95\%, 90\% and 65\%.
\label{tab:3-2-WSInt}}
\begin{tabular}{ccccc}
\hline 
Width & Width $(z=0)$ &  & \multicolumn{2}{c}{$\Sint$}\tabularnewline
(MHz) & (km/s) &  & \multicolumn{2}{c}{(mJy~MHz)}\tabularnewline
\hline 
\hline 
 &  & $C=0.95$ & $C=0.90$ & $C=0.65$\tabularnewline
\cline{3-5} 
0.22  & 46 & 0.76 & 0.49 & 0.18\tabularnewline
0.62  & 131 & 0.81 & 0.58 & 0.28\tabularnewline
1.02 & 215 & 0.87 & 0.65 & 0.35\tabularnewline
1.42  & 300 & 0.93 & 0.71 & 0.40\tabularnewline
1.82  & 384 & 0.98 & 0.77 & 0.45\tabularnewline
\hline 
\end{tabular}
\end{table}

\subsection{Reliability}

One method of estimating the reliability of an $\hi$ survey is to
use optical information provided by other large surveys, e.g. SDSS.
Another possibility is to re-observe parts of the survey area to assess
the reliability of their sources and the measured parameters, done
for example by HIPASS \citep{zwaan2004}. However, neither of these
methods are feasible for AUDS. The spectral density of SDSS in the
area of the AUDS fields is too low to systematically cross-match all
the optical and $\hi$ selected galaxies (Section \ref{sub:Optical-Counterpart})
and using additional telescope time for re-observations is not practical.

As a first step to estimate the reliability we therefore looked at
the overlap regions between the low and the high frequency bands.
The cubes overlap in the frequency range of $1368-1382\,\mathrm{MHz}$.
In this overlapping region we find 5 galaxies which are individually
detected in both the high and the low frequency bands. There are two
additional galaxies in the source list which are only detected in
the high frequency bands. However, both these galaxies have a completeness
coefficient $C<0.3.$

Next we estimated the reliability of our survey re-using the list
of possible detections created from the data cubes with inserted synthetic
objects (see Section \ref{sub:Completeness}). Of the total of 330
detected sources, 256 turned out to be synthetic, and 12 were previously
detected AUDS galaxies. Of the 62 unidentified source candidates,
about 10 passed the criteria for being included in our first cut candidate
list as described in Section \ref{sub:The-catalogue}. None of them
passed our criteria to be included in the final catalog, otherwise
they would be considered as one of the detected AUDS galaxies. We
find that the number of such false detections is roughly proportional
to the number of confirmed detections in any of the volumes. Therefore,
we conclude that first cut catalogues contain on the order of $10/256\backsimeq4\%$
false detections, and the number of false detections in the final
catalogue will be lower than that. We therefore consider 96\% as a
lower limit for the reliability of AUDS. The impact of false detections
in AUDS is negligible for all results presented in subsequent sections.

\subsection{{\normalsize{}Cosmic Variance \label{sub:3-2-5-Cosmic-Variance}}}

\begin{figure}
\begin{centering}
\includegraphics[width=1\columnwidth]{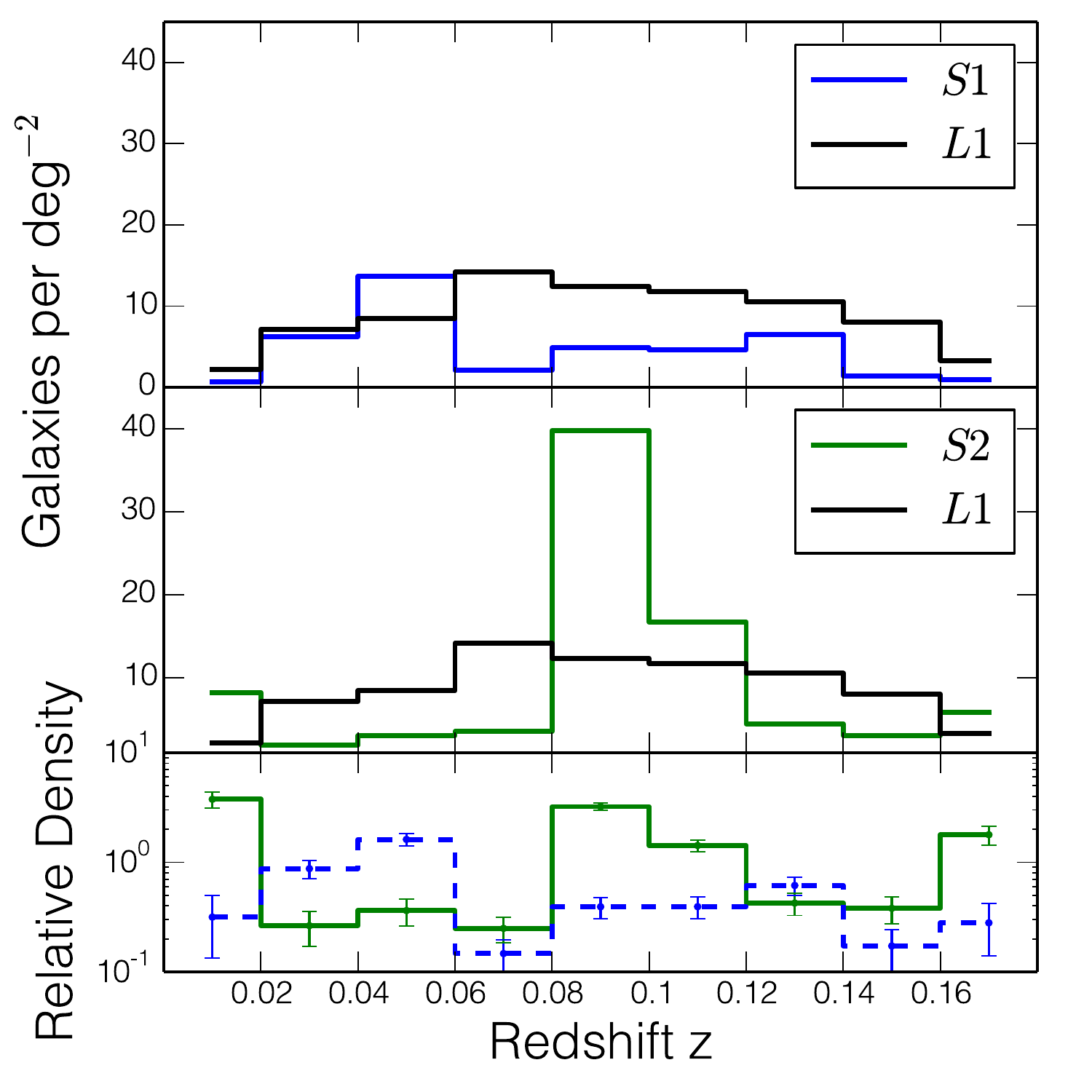}
\par\end{centering}

\protect\caption[\textcolor{black}{Comparisons between the normalised numbers of galaxies
in the $\sim4.2\,\mathrm{deg^{2}}$ sized SDSS fields S1 and S2 and
the representative SDSS field. }]{\textcolor{black}{Comparison between the normalised numbers of galaxies
in the $\sim4.2\,\mathrm{deg^{2}}$ sized SDSS fields S1 and S2 (surrounding
to two AUDS fields) and the representative SDSS field with a total
area of $5150\,\mathrm{deg^{2}}$ (L1). Top panel: Comparison between
Field 1 (blue line) and SDSS (black line), Centre panel: Comparison
between Field 2 (green line) and SDSS (black line). Bottom panel:
the relative density ratio for fields S1 and S2 compared to the representative
field L1.\label{fig:CosV} }}
\end{figure}

\begin{table}
\protect\caption[To correct our HIMF for the effect of cosmic variance multiplied we
calculated the relative density in redshift bins for both AUDS fields
individually. ]{To correct our HIMF for the effect of cosmic variance we calculated
the relative density in redshift bins for both AUDS fields individually.
We compared the number of galaxies per $\mathrm{deg}^{2}$ in redshift
bins in the small fields $(S1,S2)$ around the AUDS fields and a large
field ($L1)$ to detect over- or under-dense regions in $S1$ and
$S2$. We derive the density ratio $\rho_{\mathrm{S}}/\rho_{\mathrm{L}}$
of the AUDS field (Field~1, Field~2) in relation to a representative
SDSS galaxy sample and its dependence on the redshift.  \label{tab:CosV}
}

\centering{}%
\begin{tabular}{ccc}
\hline 
$z_{\mathrm{}}$  & $\rho_{\mathrm{S_{1}}}/\rho_{\mathrm{L1}}$  & \selectlanguage{english}%
$\rho_{\mathrm{S_{2}}}/\rho_{\mathrm{L1}}$\selectlanguage{british}%
\tabularnewline
\hline 
\hline 
$0.0-0.02$  & 0.316  & 3.754\tabularnewline
$0.02-0.04$  & 1.004  & 0.264\tabularnewline
$0.04-0.06$  & 1.835  & 0.363\tabularnewline
$0.06-0.08$  & 0.147  & 0.249\tabularnewline
$0.08-0.10$  & 0.430  & 3.216\tabularnewline
$0.10-0.12$  & 0.453  & 1.422\tabularnewline
$0.12-0.14$  & 0.439  & 0.424\tabularnewline
$0.14-0.16$  & 0.230  & 0.380\tabularnewline
$0.16-0.18$  & 0.350  & 1.783\tabularnewline
\hline 
\end{tabular}
\end{table}

Measurements of the galaxy density in a finite volume are affected
by the large scale structure of the Universe causing a bias in measurements
in small volumes like AUDS. The sample variance originating from a
finite volume is called cosmic variance. In order to quantify the
cosmic variance of the density measurement, a quantity $\xi$ can
be defined as $\xi[\%]=100\times\,\sigma_{\mathrm{var}}\,/\,\langle N\rangle$
with the variance $\sigma_{\mathrm{var}}^{2}=\Sigma_{i}(\langle N\rangle-N_{i})^{2}/n$
with $\langle N\rangle$ being the mean galaxy count in the selected
volumes, $N_{i}$ the number of galaxies in the volume $i$, and $n$
the total number of selected volumes.

To estimate $\xi$ in our sample, we selected $100$ random fields
in the SDSS North Galactic Pole field (using DR7) with the same size
and redshift range as one of the AUDS fields. We then calculated the
mean and standard deviation of the number of galaxies in these fields.
To estimate the sampling error in $\xi$ and $\sigma_{\mathrm{var}}$
we repeated that procedure 1000~times. The number of galaxies per
random field varied between 10 and 335 galaxies with an average number
of $54(\pm2)\,\pm\,23(\pm3)$ galaxies or a sample variance of $\xi=42\pm5$
\%. Doing the same test, selecting two AUDS sized fields as in the
survey reduces the cosmic variance to $\xi=29\pm4\%$. The decrease
in cosmic variance corresponds to a reduction of $\sqrt{2}$. Comparing
these results to the information from our fields, we find that our
fields are slightly denser than the average field of the same size
in SDSS. We find 58 galaxies in Field 1 (1.1 times over-dense) 60
galaxies (1.2 times over-dense) in Field 2.

We compare our results to those presented by \citet{driver2010}.
They derived an empirical expression for $\xi$ for galaxy surveys
using the spectroscopic redshifts of galaxies from the DR7 SDSS in
the redshift range of $0.03<z<0.1$. They found that $\xi$ is mainly
influenced by the survey volume, the survey aspect ratio and whether
the survey area is contiguous or consists of several independent volumes.
Using their equation we find $\xi=33\%$ for both fields, or $ $$\xi=46\%$
which is within $1\sigma$ of our result.

The sample variation in the AUDS fields suggest that any density result
will be correspondingly biased. However, a correction is possible
as follows: 

(1) We selected a large sub-sample from the spectroscopic DR7 SDSS
galaxy sample ($L1$). $L1$ is in the main SDSS field ($130^{\mathrm{o}}<\alpha<236^{\mathrm{{o}}}$,
$0^{\mathrm{o}}<\delta<58^{\mathrm{o}})$ selected to maximise the
area (5150 deg$^{2}$) but to avoid the complex shape of the edges
of SDSS (see Driver \& Robotham 2010). Even though SDSS is not completely
immune to cosmic variance itself, $L1$ is large enough that the expected
difference between the mean density of $L1$ and that of the Universe
is $\sim7\%$ based on the results of \citet{driver2010}.

(2) We selected all SDSS galaxies in small fields surrounding each
of the AUDS fields. We name these smaller SDSS fields $S1$ and $S2$.
The area of $S1$ and $S2$ needs to be larger than the area of the
original AUDS fields to reduce Poisson noise due to the small number
of galaxy counts, but small enough such that their density remains
correlated with the density of galaxies in the AUDS fields.

To find the optimum size of $S1$ and $S2$, we placed 100 AUDS-like
volumes at random positions within SDSS. We then computed the average
galaxy density of each AUDS-like volume. Around each of these 100
AUDS like fields we placed another field of a larger size. For each
field, we calculated the ratio of the density of the AUDS sized field
to the larger field surrounding it. Repeating this procedure for differently
sized field 250 times, we found that the standard deviation of the
density ratio is minimised (at 11\%) when the size of the surrounding
field is 4.2 deg$^{2}$. That is, the quadrature sum of Poisson noise
and cosmic variance are lowest for this field size. Furthermore, the
density ratio itself indicates that $S1$ and $S2$ are representative
of the structure in the smaller field.

(3) We compared the number of galaxies per area in redshift bins in
the small fields $(S1,S2)$ and the large field ($L1)$ to detect
over- or under-dense regions in $S1$ and $S2$ (Figure \ref{fig:CosV},
Table \ref{tab:CosV}). It is important to note that we make the assumption
that the optically selected fields ($S1$,$S2$ and $L1)$ have the
same distribution of galaxies as the $\hi$ selected AUDS sample.
Looking at the respective redshift bins for each field we find the
relative density of the small fields to vary between being 4.3 times
under-dense and 3.8 times over-dense in comparison to the representative
SDSS field ($L1$). Tracing the relative density in the AUDS fields
in redshift bins allows us to correct for the effect in the HIMF as
described in Section \ref{sub:HIMF-Methods}. 

For this correction we make the assumption that optical data and the
$\hi$ data correlate. To test this assumption we compare the bias
factor for the SDSS fields $(S1,\, S2,\, L1)$ with derived bias parameters
for $\hi$ selected surveys. \cite{Seljak} measured the bias parameter
for optical galaxies as a function of luminosity. Using their bias
factors we find that the average bias factor of the galaxies in the
4.2 deg$^{2}$ regions around the AUDS fields are: \foreignlanguage{english}{$\langle b_{S1,S2}\rangle=0.97\pm0.11$
while the $L1$ field has an average bias of }$\langle b_{L1}\rangle=0.99\pm0.06$.
Observations as well as numerical simulations estimate local bias
parameters of the neutral hydrogen relativ to the dark matter between
0.7 and 1.0, with a typical uncertainty of$\pm0.2$ (Basilakos et
al. 2007, Martin et al. 2012, Dav\'e et al. 2013, see also Padmanabhan
et al. 2015 for review). This shows that the clustering between optical
selected galaxies and dark matter, and $\hi$ selected galaxies and
dark matter are very similar on the spatial scales probed here, therefore
allow to use the optical data to correct for over-/under-densities
in the $\hi$ selected galaxies.

\section{$\hi$ Mass function (HIMF) \label{sec:3-3Mass-function}}

\subsection{Methods\label{sub:HIMF-Methods}}

\begin{table*}
\protect\caption{Comparison of different results for the HIMF from 21~cm surveys.
\label{tab:Comparison-HIMF}}

\begin{tabular}{ccccc}
\hline 
Survey  & Reference  & $\alpha$  & $\log\,(M_{\hi}^{*}/M_{\odot}$)  & $\Phi^{*}$ \tabularnewline
 &  &  & $+2\log h$  & $(10^{-3}\, h^{3}\mathrm{Mpc^{-3})}$\tabularnewline
\hline 
AHISS  & {\footnotesize{}\citet{Zwaan1997} } & $-1.2$  & 9.55  & 14\tabularnewline
ADBS  & {\footnotesize{}\citet{Rosenberg2002} } & $-1.53$  & 9.63  & 11.9\tabularnewline
 & {\footnotesize{}\citet{Springob2005} } & $-1.24$  & 9.68  & 9.3\tabularnewline
HIPASS  & {\footnotesize{}\citet{Zwaan2005} } & $-1.37\pm0.03$  & $9.55\pm0.02$  & $14.2\pm1.9$\tabularnewline
ALFALFA  & {\footnotesize{}\citet{Martin2010} } & $-1.33\pm0.02$  & $9.71\pm0.01$  & $14.0\pm0.9$\tabularnewline
AUDS  & \textbf{This work } & $\alC$  & $\MiC$  & $\PLC$\tabularnewline
\hline 
\end{tabular}
\end{table*}

The HIMF $\Phi(\MHI)$ is a measure of the number of galaxies per
unit volume $\mathrm{d}V$for a given $\MHI$ and is crucial input
parameter for models and simulations describing galaxy formation and
evolution. We derived the HIMF in co-moving coordinates to avoid changes
in the measured densities purely caused by the expansion of the Universe.

The HIMF is often parameterised by a Schechter function defined as:

\begin{equation}
\Phi(M_{\mathrm{{\hi}}})=\ln10\,\Phi^{*}\,\bigg(\frac{{M_{\mathrm{\hi}}}}{M_{*}}\bigg)^{\alpha+1}e^{-\frac{M_{\hi}}{M_{*}}},\label{eq:Schechter-1-1}
\end{equation}
with the faint end slope $\alpha$, the characteristic mass $M_{\mathrm{\hi}}^{*}$
and the normalisation $\Phi^{*}$.

We use two different methods to derive the HIMF: The $\sum\,1/\Vmax$
method \citep{schmidth1968} and the 2D stepwise maximum likelihood
(SWML) \citep{Zwaan2003}.

The basic $1/\Vmax$ method assigns each galaxy a weighting factor
which corresponds to the inverse maximum volume $(\Vmax)$ in which
a galaxy can be detected inside the survey volume. 

We adapt this method to compute the maximum search volume using the
relation for completeness in equation \ref{eq:Cross2-1-1}. The completeness
for a galaxy changes if the galaxy is shifted to a different part
of the cube as the noise changes within the field and with frequency.

We therefore create two additional data cubes of the same size as
the original cubes. In the first one, each pixel value corresponds
to the co-moving volume in $\mathrm{Mpc}^{3}$ corresponding to that
pixel. In the second data cube each pixel value corresponds to the
RMS noise of neighbouring pixels. The noise is computed by calculating
the RMS noise in a $\pm250\,\mathrm{km/s}$ range around that pixel.
For each galaxy, we computed the expected completeness $C_{i}$ if
that galaxy were placed at any of the pixels within the survey volume
using equation \ref{eq:Cross2-1-1}. For that purpose, we scaled $\Sint$
and $W$ to the distance corresponding to each pixel, and used the
RMS of that pixel in the noise cube. The effective volume per pixel
is then the product of completeness and volume $V_{i}$ for each galaxy.

The volume in which a given galaxy is detectable, hereafter called
the ``detectable volume'' is then the sum over the effective volume
of each pixel 
\begin{equation}
V_{\mathrm{{max}}}=\sum_{i}C_{i}\times V_{i}.\label{eq:maxV}
\end{equation}
The co-moving HIMF is then defined as the sum over the $\MHI$ range
of all galaxies $j$ in a $\MHI$ bin with the bin size $\Delta\MHI$
\begin{equation}
\Phi(\MHI)=\sum_{j}\frac{{1}}{V_{\mathrm{max},j}}.
\end{equation}

In addition to the volume cube, we also created a cosmic-variance-corrected
volume cube. For the cosmic-variance-corrected cube we multiplied
each volume pixel in a redshift bin with the relative density of that
redshift bin found in the optical sample (Table: \ref{tab:CosV})

\begin{equation}
\Vmaxc[z_{1},z_{2}]=\frac{\rho_{S}[z_{1},z_{2}]}{\rho_{L}[z_{1},z]}\Vmax[z_{1},z_{2}].
\end{equation}
That means that the volume of a pixel in an under-dense region is
``shrunk'' while the volume of a pixel in an over-dense regions
is ``enlarged''. The $\sum1/\Vmax$ method has the advantage of
being fast and simple to implement as well as producing a normalised
HIMF.

The second method to calculate the HIMF is the stepwise maximum likelihood
method developed by \citet{Efstathiou1988} as a superior tool to
derive the luminosity functions of galaxies. The idea behind the SWML
technique is to find the function $\Phi(\MHI)$ that yields maximal
joint probability of detecting all galaxies in the sample. It has
the advantage over the $1/\Vmax$ method because the results are independent
variables of density. In the SWML method the galaxy mass distribution
is split into bins assuming a constant distribution in each bin. It
is not necessary to assume a functional form for the HIMF. Based on
SWML method, \citet{Zwaan2003} developed the 2 dimensional - SWML
method which solves for the space density of $\MHI$ and $W$ at the
same time.

\subsection{{\normalsize{}$\sum\,1/\Vmax\,$ HIMF\label{sub:-HIMF-Vmax}}}

\begin{figure}
\begin{centering}
\includegraphics[width=1\columnwidth]{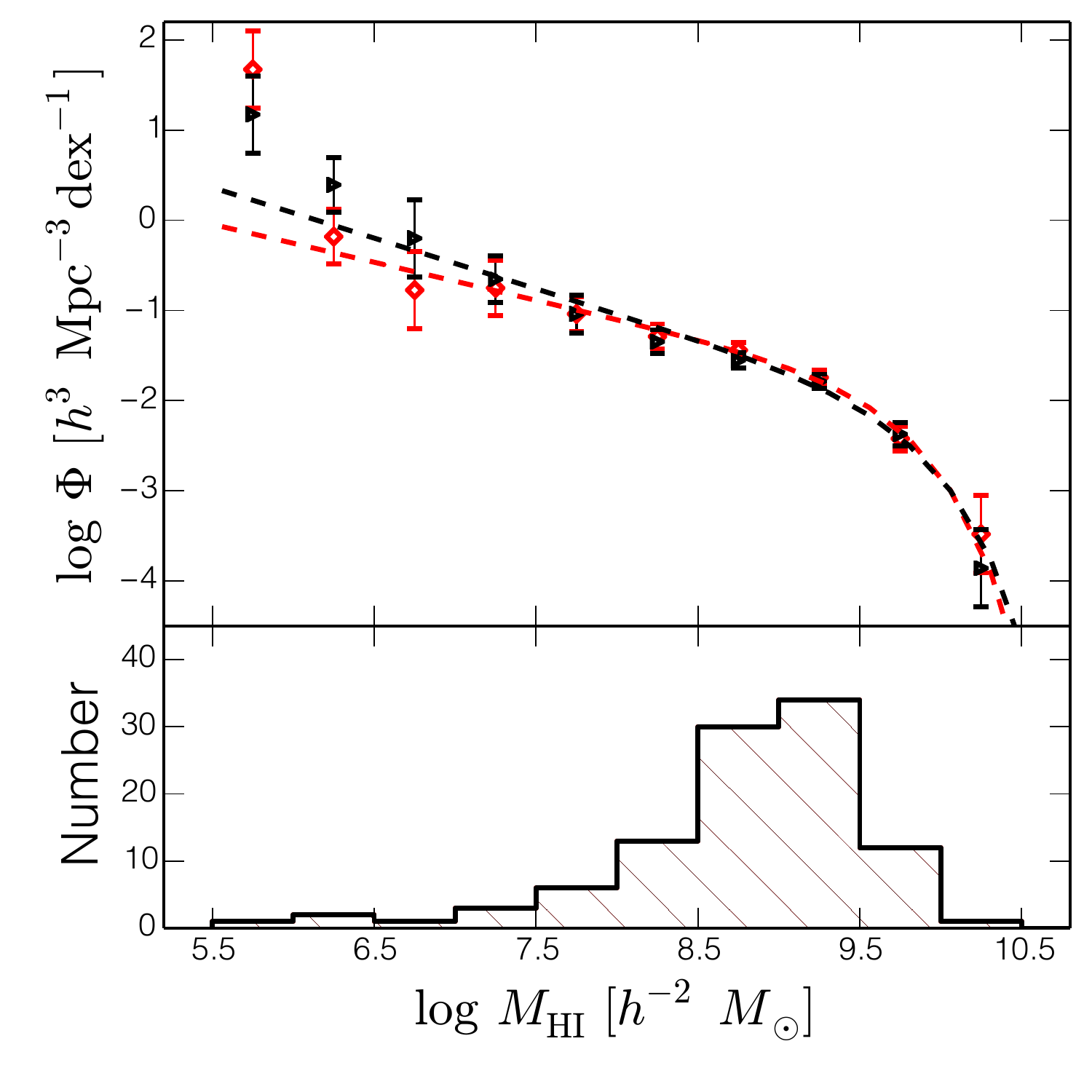}
\par\end{centering}

\protect\protect\caption[The AUDS co-moving $\hi$ mass function measured using the $\Sigma\,1/\Vmax$
method and $\Sigma1/\Vmaxc$ method (top panel) and Histogram of the
$\hi$ masses (lower panel). ]{Top panel: The AUDS co-moving $\hi$ mass function measured using
the $\Sigma\,1/\Vmax$ method (black triangles) and $\Sigma1/\Vmaxc$
method (red diamonds), with their corresponding best-fit Schechter
functions. The errorbars indicate the 1$\sigma$ uncertainties based
on Poisson statistics. The black dashed line is the best-fit Schechter
function for the $\Sigma\,1/\Vmax$ with $\alpha=\al$, $\Phi^{*}=(\PL)\,\times10^{-3}\, h^{3}\,\mathrm{Mpc^{-3}}$
and $\log(\MHI^{*}/M_{\odot})=(\Mi)+2\log h$. The red dash-dotted
line is the best-fit Schechter function for the $\Sigma\,1/\Vmaxc$
method with $\alpha=\alC$, $\Phi^{*}=(\PLC)\times10^{-3}\, h^{3}\,\mathrm{Mpc^{-3}}$
and $\log(\MHI^{*}/M_{\odot})=(\MiC)+2\log h$. Lower panel: Histogram
of $\hi$ masses, using the same binning as in the top panel.\label{fig:HIMF1}
}
\end{figure}

Figure \ref{fig:HIMF1} shows the results of the HIMF calculation
using the $\sum\,1/\Vmax$ method. In the following we will exclude
the lowest mass data point in our fit as it only contains one galaxy
with a low completeness coefficient (see Figure \ref{fig:AUDS_spec}
upper left panel). Nevertheless, this galaxy and the two in the next
bin are suggestive of a possible upturn in the HIMF .

The errorbars were calculated using Poisson statistics not taking
into account measurement errors (e.g. distance and $\MHI$). The best
Schechter fit to the model without correction for cosmic variance
is $\alpha=\al$, \foreignlanguage{english}{$\log(\MHI^{*}/M_{\odot})=(\PL)+2\log h$}
and $\Phi^{*}=(\PL)\,\times10^{-3}\, h^{-3}\,\mathrm{Mpc^{-3}}$.
The errors in the Schechter function are computed using a jackknife
technique, de-selecting one galaxy at a time \citep{Quenouille1949PCPS...45..483Q}.

Next we use the cosmic-variance-corrected volumes ($\Vmaxc$) to compute
the HIMF (Figure \ref{fig:HIMF1}). Both measurements show good agreement
within the errorbars. The result for the best-fit to the cosmic variance
corrected data is $\alpha=\alC$, $\log(\MHI^{*}/M_{\odot})=(\MiC)+2\log h$
and $\Phi^{*}=(\PLC)\times10^{-3}\, h^{3}\,\mathrm{Mpc^{-3}}$.

We binned the data in 0.5~dex bins of $\MHI$ starting at an $\hi$
mass of 5.5. Different binning and starting masses give slightly different
mass functions and fits. We estimated the error caused by this by
varying the starting mass in the range of $\log(\MHI/M_{\odot})=(5-5.5)+2\log h$
and the bin size by 2-5 bins per dex in $\MHI$. The RMS over the
best-fits was comparable to the fitting error ($\sigma_{\alpha}=0.04$,
$\sigma_{\log\MHI^{*}}=0.10$ and $\sigma_{\Phi}=4.6\times10^{-3}\, h^{3}\,\mathrm{Mpc}^{-3}$
).

\subsection{2DSWML HIMF}

\begin{figure}
\centering{}\includegraphics[width=1\columnwidth]{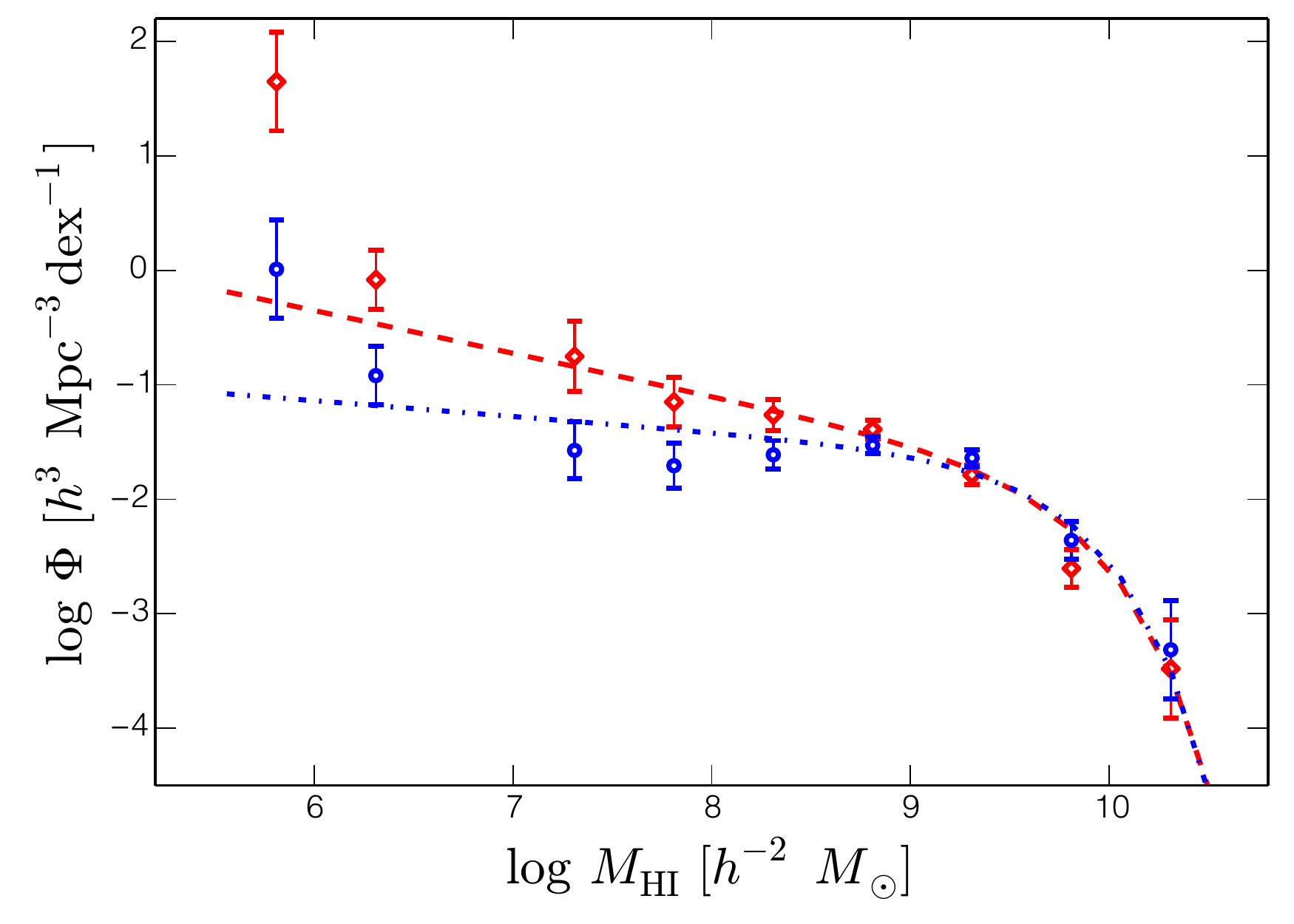}\protect\caption[Comparison between the HI mass functions derived from the $\Sigma1/\Vmaxc$
method and from the 2DSWL method.]{Comparison between the HI mass functions derived with the $\Sigma1/\Vmaxc$
method (red diamonds and dashed line) and with the 2DSWL method (blue
dots and dash- dotted line, respectively). The lines indicate the
best-fit Schechter functions.\label{fig:HIMF_2d-1} }
\end{figure}

The 2DSWML method is independent of density variations due to large
scale structure and therefore a good, independent test to check the
results from the $\Sigma\,1/\Vmax$ method. It does, however, rely
on the shape of the HIMF not changing. Furthermore, the difference
in the noise at different locations within the survey volume cannot
be taken into account. Instead, we used a single value for the RMS
for each detected galaxy.

We recovered the normalisation for the HIMF using the mean galaxy
density $\bar{n}$. $\bar{n}$ is calculated by correcting the measured
distribution of galaxies with the selection function $S(D)$ (Section
\ref{sub:Selection-function}). \citet{Davis1982} presented several
estimators to calculate $\bar{n}$. We choose the $\overline{n_{3}}=N_{\mathrm{total}}/\int S(D)\,\mathrm{d}V$
estimator as it is the most stable one for small numbers even though
it has a slight dependency on large scale structure.

Figure \ref{fig:HIMF_2d-1} compares the 2DSWML HIMF and the $\Sigma\,1/\Vmaxc$
HIMF. We find that the data points for both methods are in reasonably
good agreement at the high mass end ($\MHI>10^{9.5}M_{\odot})$ with
each other. At lower masses the slope of the 2DSWML (red dashed line)
is less steep. Even though most of the data points agree with each
other within the $1\sigma$ errorbar the 2DSWML HIMF points are systematically
lower causing a significant difference in the fitted slope. The best-fit
to the 2DSWML HIMF function yields $\Phi^{*}=(\PLSWML)\times10^{-3}\, h^{3}\,\mathrm{Mpc}^{-3}$,
$\log(\MHI^{*}/M_{\odot})=(\MiSWML)+2\log h$ and $\alpha=\alSWML$.
The good agreement between the normalisation of the 2DSWML and the
$\Sigma\,1/\Vmaxc$ HIMF is encouraging, as the normalisation for
the 2DSWML HIMF was calculated independently.

\subsubsection{\label{sub:Selection-function}Selection Function }

The selection function $S(D)$ is the probability that a galaxy at
a distance $D$ is detected by the survey. We calculate $S(D)$ for
the 2DSWML and the $\sum1/\Vmax$ HIMF as described in \citet{Zwaan2003}.

Assuming a homogeneous space distribution of galaxies, the number
of galaxies observed in a distance bin of the size $\Delta D$ at
the distance D is $n(D)=\Omega\, D\Delta D\:\overline{n}\, S(D)$
with the solid angle $\Omega$ and the average number of galaxies
$\bar{n}$. Figure \ref{fig:MHI_hist} compares the detected and the
predicted redshift distribution showing that they are not in good
agreement.

The galaxy numbers derived by the selection function give lower numbers
of nearby galaxies while overestimating galaxies at larger distances
in comparison to the detected galaxy distribution. The effects could
be caused by large scale structure which are not traced by the selection
function. A good example is the over-density at $D\approx270\mathrm{\,\mathrm{Mpc}}$
in the AUDS histogram which does not show up in the prediction. A
possible explanation for the over-estimation of high redshift galaxies
lies in the general limitation of AUDS to pick up galaxies at high
redshift largely caused by RFI, as described in Section \ref{sub:The-catalogue}.

\begin{table*}
\protect\caption{Comparison between the best Schechter fit as well as the values for
$\OMHI$ for the derived AUDS HIMF. }

\begin{tabular}{cccccc}
\hline 
Methods  & $\alpha$  & $\log\,(M_{\hi}^{*}/M_{\odot}$)  & $\Phi^{*}$  & $\Omega_{\hi}$ (int.)  & $\Omega_{\hi}$ (sum.)\tabularnewline
 &  & $+2\log h$  & $(10^{-3}\, h^{3}\,\mathrm{Mpc}^{-3})$  & $(10^{-4}\, h^{-1})$  & $(10^{-4}\, h^{-1})$\tabularnewline
\hline 
\hline 
$\Sigma\,1/\Vmax$  & $ $$\aa$$\al$  & $\Mi$  & $\PL$  & $\OI$  & $\OS$\tabularnewline
$\Sigma\,1/\Vmaxc$  & $\alC$  & $\MiC$  & $\PLC$  & $\OIC$  & $\OSC$\tabularnewline
2DSWML  & $\alSWML$  & $\MiSWML$  & $\PLSWML$  & $\OISWML$  & $\OSSWML$$ $\tabularnewline
\hline 
\end{tabular}
\end{table*}

\subsection{Influence of Completeness}

\begin{figure*}
\begin{centering}
\includegraphics[bb=10bp 0bp 288bp 432bp,height=6cm]{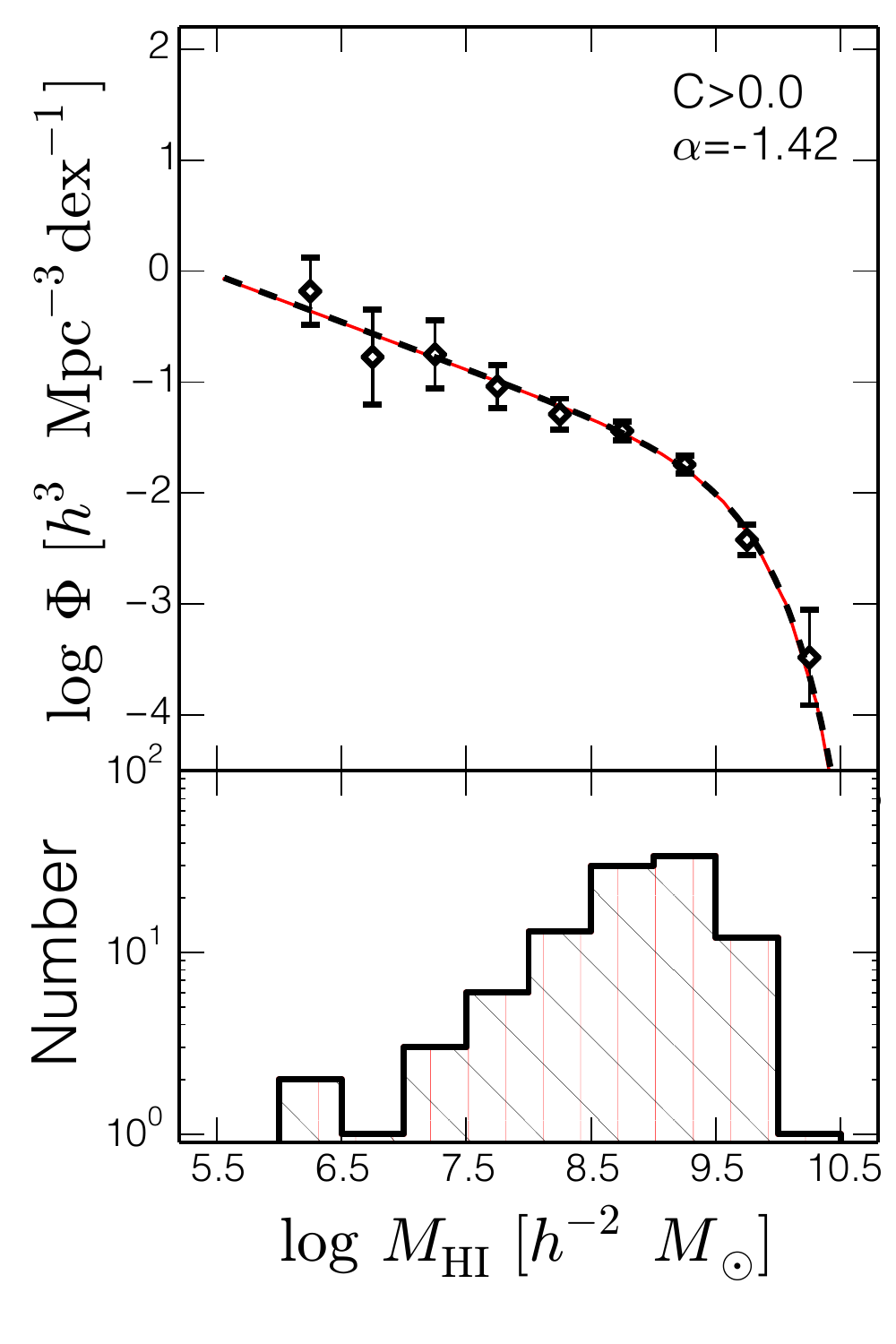}\includegraphics[bb=56bp 0bp 288bp 432bp,clip,height=6cm]{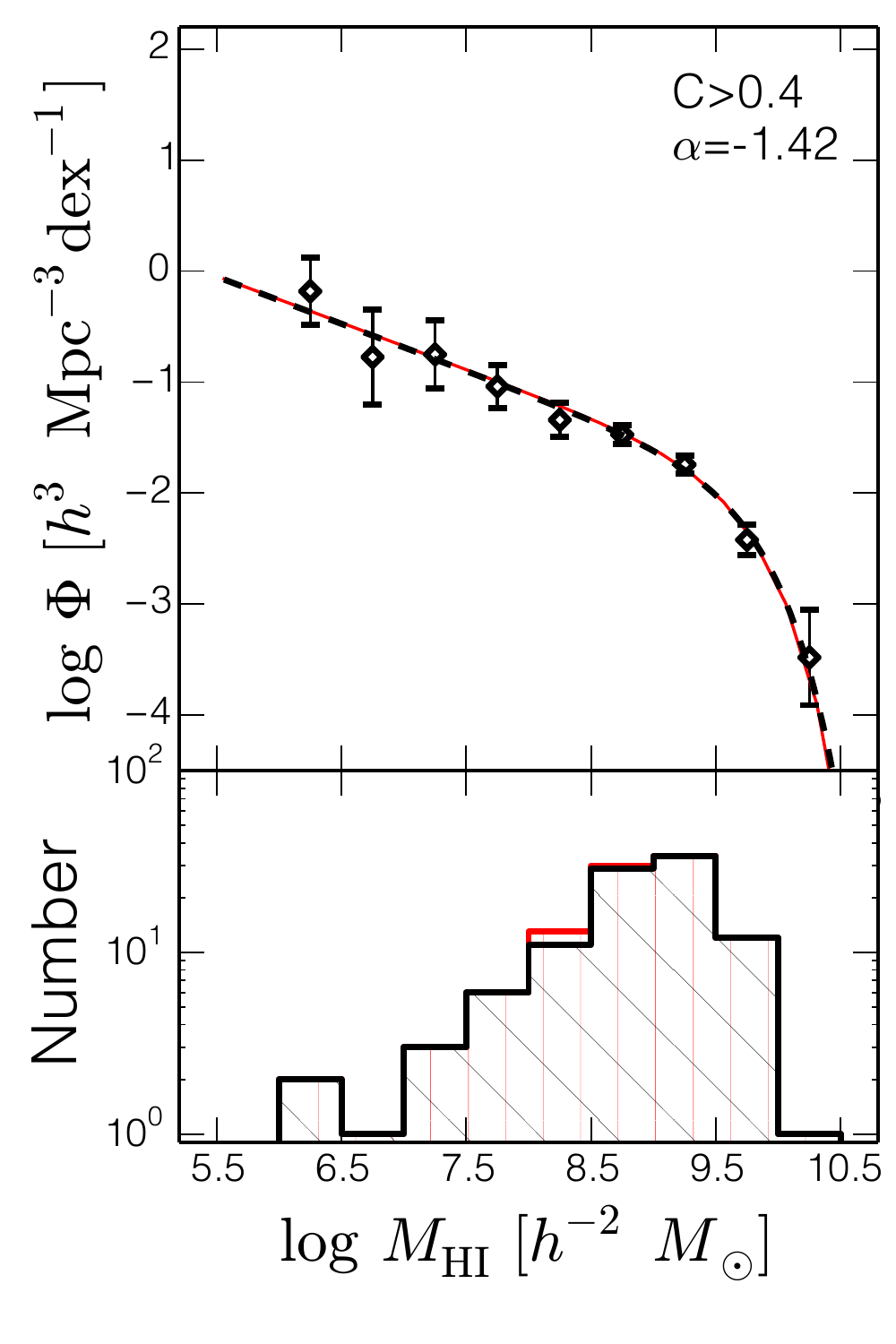}\includegraphics[bb=56bp 0bp 288bp 432bp,clip,height=6cm]{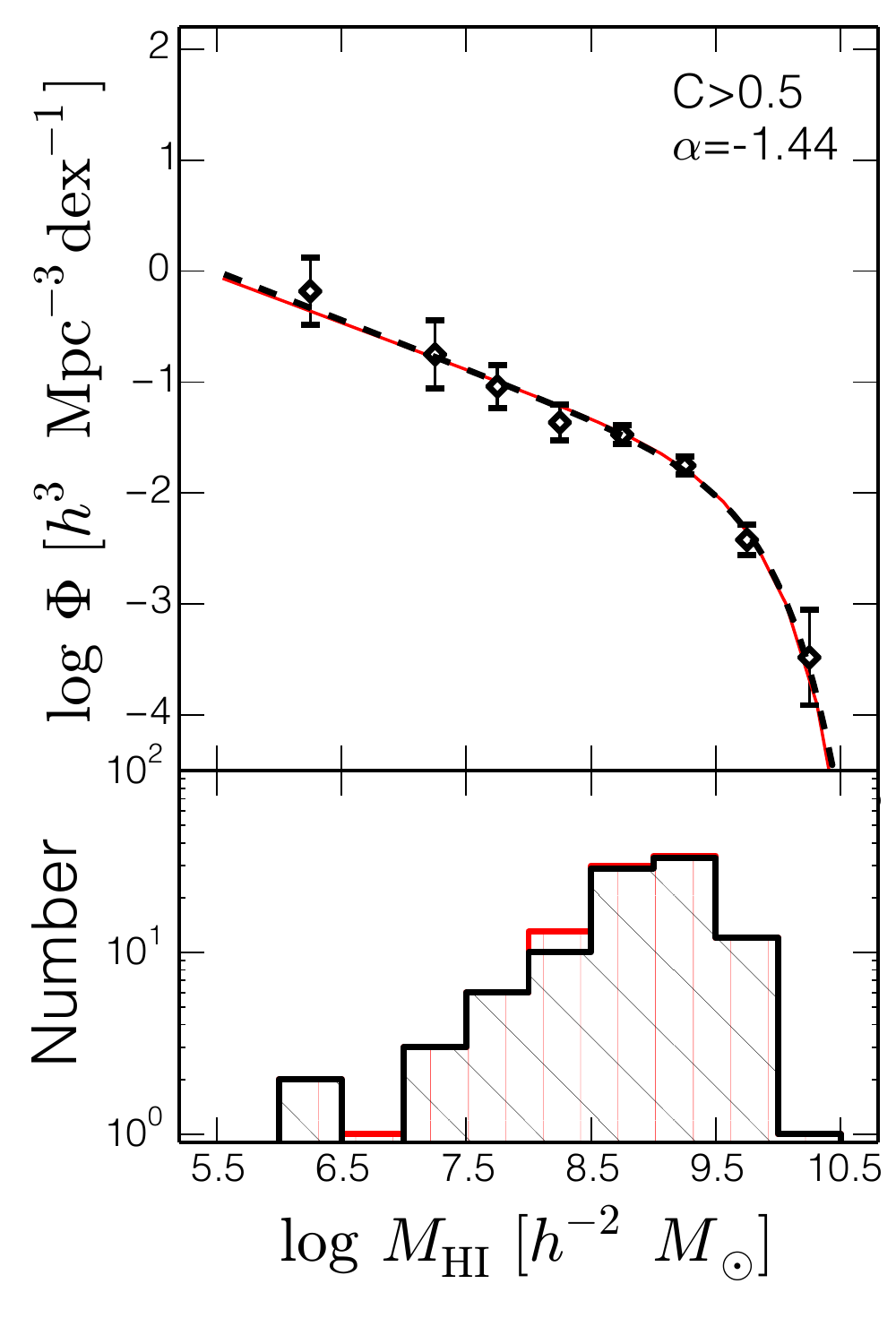}\includegraphics[bb=56bp 0bp 288bp 432bp,clip,height=6cm]{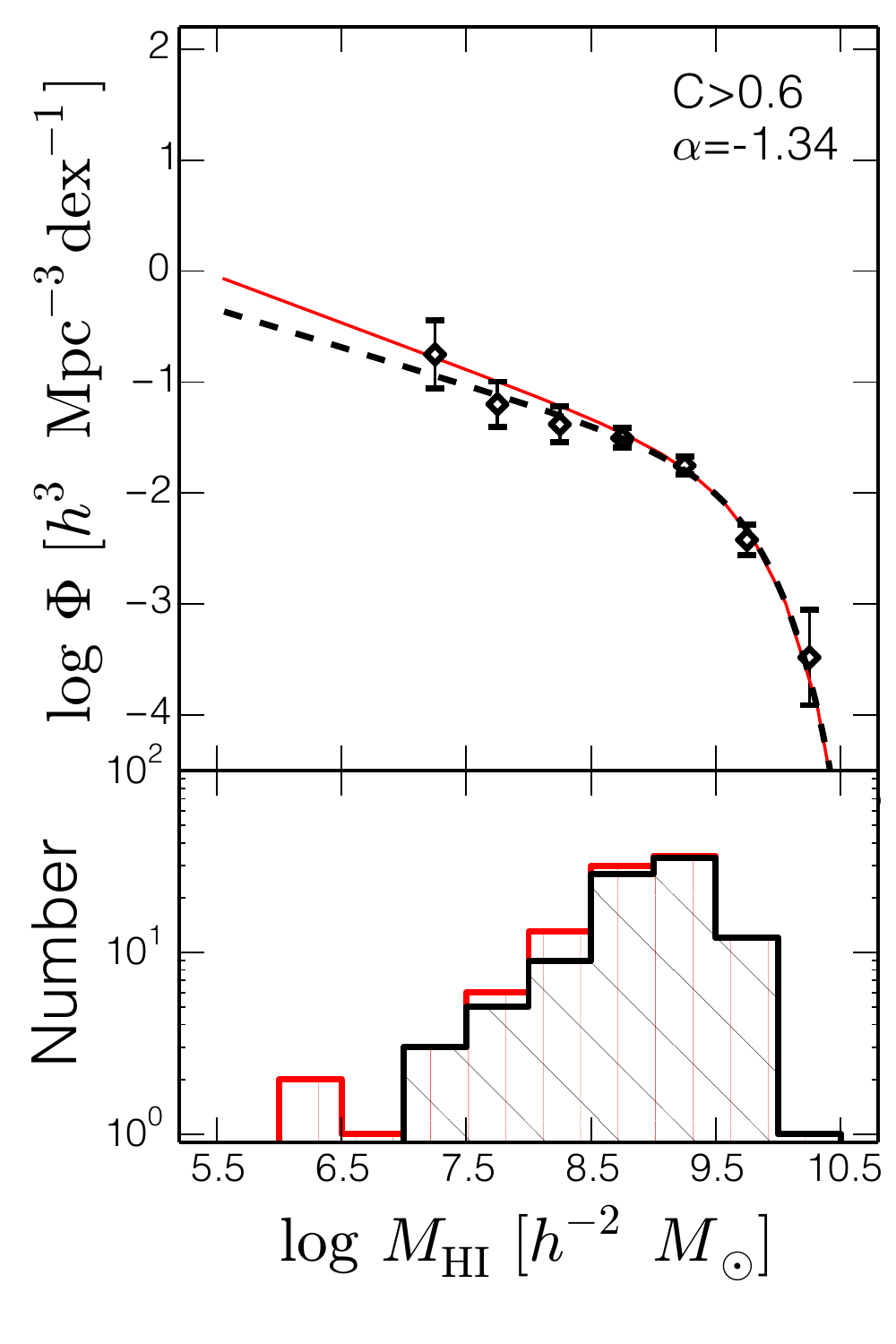}
\par\end{centering}

\protect\protect\caption[The \foreignlanguage{english}{$1/\Vmaxc$} HIMF calculated for galaxies
selected with different completeness cut offs.]{The \foreignlanguage{english}{$1/\Vmaxc$} HIMF calculated for galaxies
selected with different completeness cutoffs. The completeness equation
(\ref{eq:Cross2-1-1}) was used to measure the completeness coefficient
$C$ for each galaxy. Plots for $C>0,\,0.4\,,0.5$ and $0.6$ are
shown. The corresponding numbers of galaxies are 102, 98, 96 and 92.
The Schechter fits are shown as the black dotted lines with the result
of the fitted slope in the top right corner of each panel. The red
solid line is the Schechter fit from Figure \ref{fig:HIMF1}, based
on all galaxies with $C>0$. The red histograms show the HI mass distribution
of the full AUDS sample and the black histograms are the corresponding
distributions for the restricted samples. \label{fig:Completeness-Cut}
 }
\end{figure*}

We rate the quality of our detected galaxies by their completeness
coefficient C and only include galaxies with $C>0$, excluding the
lowest mass galaxy for the HIMF fit. In Figure \ref{fig:Completeness-Cut}
we compare the $1/\Vmaxc$ HIMF for different cuts in the completeness
coefficient ($C=0.4\,,0.5,\,0.6)$. The plot shows the lower mass
galaxies ($M<10^{9.5}M_{\odot})$ are excluded first causing the slope
to flatten. High completeness cut-offs $(C>0.5)$ exclude galaxies
with masses around the knee of the HIMF resulting in smaller values
for the normalisation $\Phi^{*}$. The change in the HIMF with completeness
is an intriguing result as unlike HIPASS and ALFALFA, the low-mass
AUDS galaxies are located well beyond the local volume and may represent
more typical volumes in the Universe.

\subsection{Evolution of the HIMF}

\begin{figure}
\begin{centering}
\includegraphics[width=1\columnwidth]{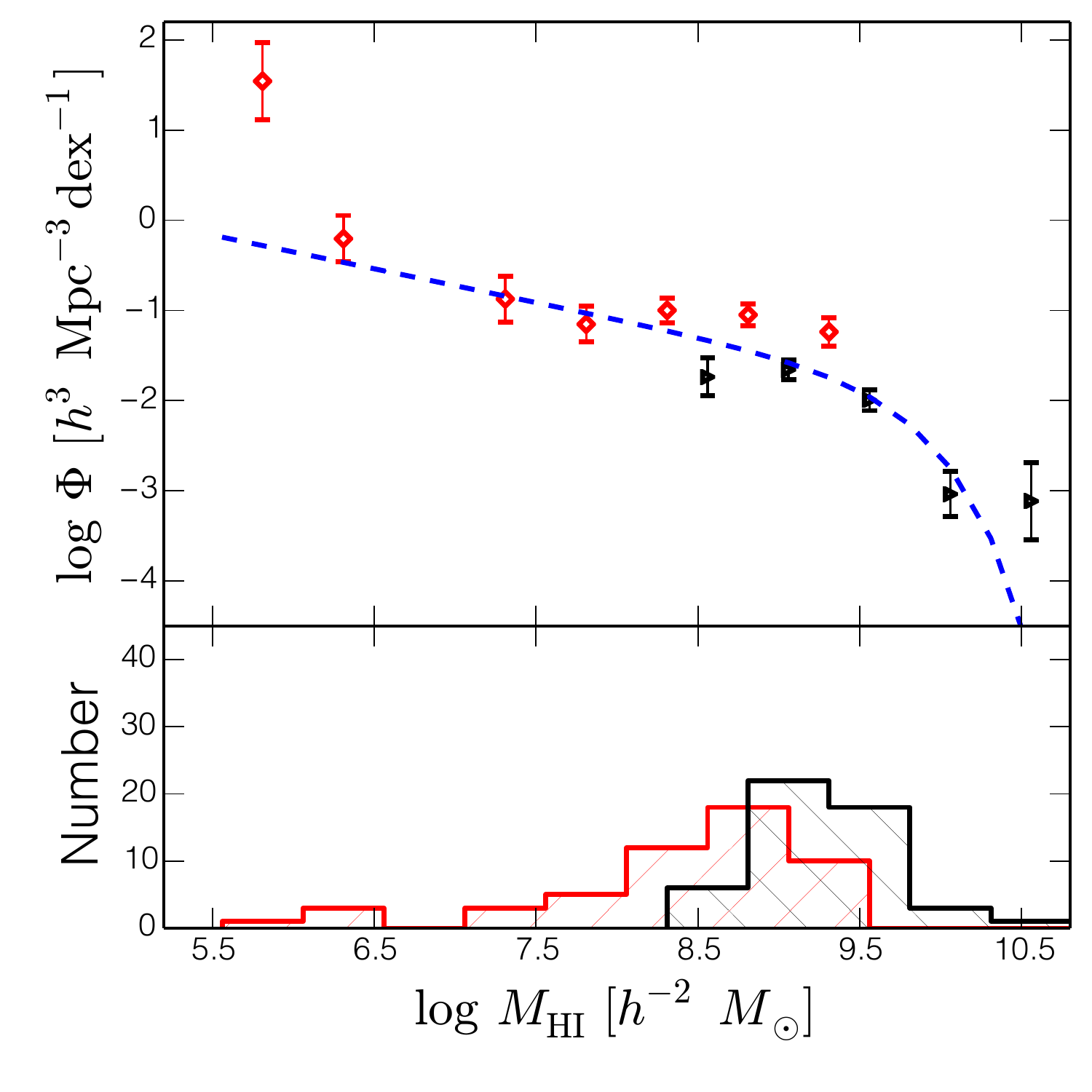}
\par\end{centering}

\protect\protect\caption[Comparison between the HIMF derived from the low redshift and high
redshift sub-sample (top panel) and the number distribution of the
low and high redshift sample. ]{Top panel: Comparison between the HIMF derived from the low redshift
(red diamonds) and high redshift (black triangles) sub-sample. The
blue dashed line is the fit to the complete AUDS sample (Figure 11).
In the mass range of $10^{8}-10^{10}h^{-2}M_{\odot}$ the low redshift
HIMF is significantly higher than the high redshift HIMF.  The discrepant
high redshift bin at $10^{6.75}h^{-2}M_{\odot}$ only contains one
galaxy. Lower panel: The number distribution of the low redshift sample
$(\langle z\rangle=0.036)$ in red and the high redshift sample $(\langle z\rangle=0.094)$
in black. \label{fig:HIMFz} }
\end{figure}

The redshift range of AUDS allows us to split the sample into redshift
bins to see evolutionary trends. We split our sample at the mean redshift
of our sample $(z=0.065)$. This creates a lower redshift bin with
52 galaxies and a mean redshift of $z=0.036$ and higher redshift
bin with 50 galaxies and a mean redshift of $z=0.095$. The binning
by redshift indirectly means we divide the galaxies by their mass,
as the faint galaxies can only be detected close by while rarer, massive
galaxies need the large volume of the high redshift bin to be found.

However, there is some overlap over a mass range of $10^{8}$ to $10^{10}M_{\odot}$
as can be seen in Figure \ref{fig:HIMFz}. There appears to be evidence
for only a modest change in the HIMF. Formally, if $\alpha$ and $\MHI^{*}$
are held fixed at their best-fit value for the whole sample, $\Phi^{*}$
is $\sim16\%$ lower for the higher redshift sample. Any change in
the density of galaxies with $\hi$ mass $>10^{10}$ or $<10^{8}$
$M_{\odot}$ cannot be explored. The full AUDS sample will provide
tighter limits on the amount of evolution that can arise from feedback
processes in galaxies over this redshift range \citep{Kim2013}.

\subsection{Comparison of HIMF}

The HIMF is a useful tool to describe how much $\hi$ is locked up
in galaxies. The slope of the HIMF gives the relative importance of
low mass and high mass galaxies. An early HIMF measured from the AHISS
survey \citep{Zwaan1997} found a flat faint end slope of $\alpha=-1.2$.
The Arecibo Dual-Beam Survey (ADBS) \citep{Rosenberg2002} found a
much steeper slope of $\alpha=-1.53$. However, both these surveys
suffer from small numbers and small volumes.

\begin{figure}
\begin{centering}
\includegraphics[width=1\columnwidth]{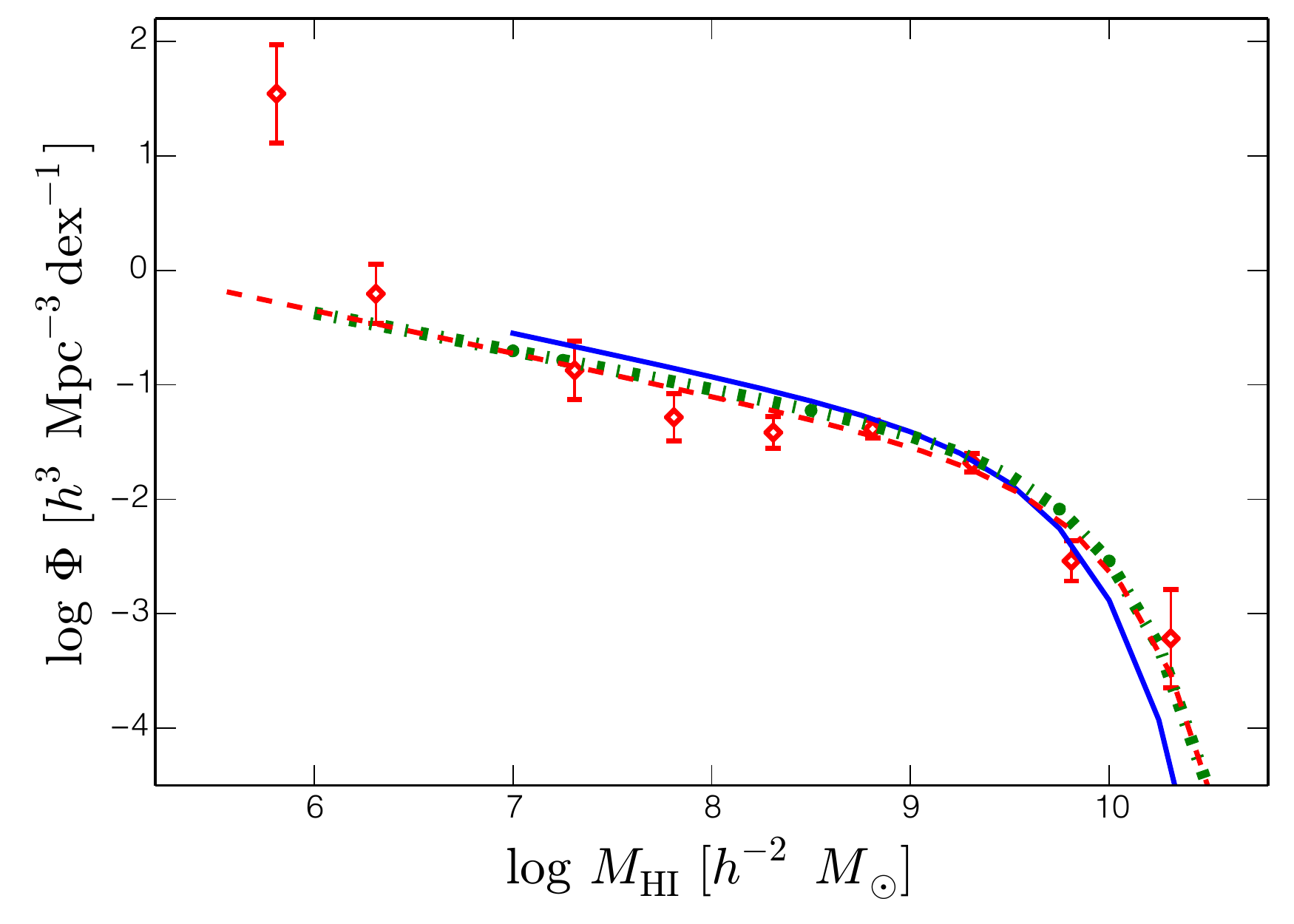}
\par\end{centering}

\protect\caption[Comparison between the AUDS HIMF and the best-fit Schechter functions
from HIPASS and ALFALFA. ]{Comparison between the AUDS HIMF (red diamonds and red dashed line)
and the best-fit Schechter functions from HIPASS (blue solid line)
and ALFALFA (green dot dashed line). The individual data points of
the HIMF agree (within the errorbars) with the two surveys in the
local Universe, apart from a small number of low $\hi$ mass AUDS
galaxies. \label{fig:HIMF_Com} }
\end{figure}

The two largest blind $\hi$ surveys HIPASS and ALFALFA (40\%) find
slopes $\alpha=-1.37\pm0.03$ \citep{Zwaan2005} and $\alpha=-1.33\pm0.02$
\citep{Martin2010} respectively. The good agreement between the two
surveys seems to suggest that the slope of the HIMF in the local Universe
is well defined. However, at the high mass end the ALFALFA survey
reveals a larger number of galaxies than HIPASS (Figure \ref{fig:HIMF_Com}).
\citet{Martin2010} explains this difference with the higher upper
redshift limit and larger volume in comparison to HIPASS.

The AUDS sample allows us to construct an HIMF for the first time
which is independent of the local volume and at much higher redshifts
and to higher sensitivities than previous surveys. Figure \ref{fig:HIMF_Com}
compares the results from AUDS with the best-fit of HIPASS and ALFALFA.
AUDS measures a slightly steeper slope $\alpha$ than HIPASS or ALFALFA,
but overall the surveys agree well with each other. We find a tentative
rise at very low masses caused by faint, low-mass galaxies detected
in our survey, which might have been missed in previous, less sensitive
surveys. Twice as many galaxies are detected compared to the prediction
from the extrapolated HIMF. Unfortunately the overall small-number
statistics make it necessary to interpret this result with caution.
Although the detected faint galaxies are well beyond the Local Group,
the volume sampled at this mass level is only about $8\,\, h^{-3}\,\mathrm{Mpc}^{3}$,
so the cosmic variance is high.

\section{{\normalsize{}Cosmic $\hi$ Density $\Omega_{\hi}$ \label{sec:3-4Cosmic--Density}}}

\subsection{{\normalsize{}$\hi$ Mass Density$(\rho_{\hi})$ }\foreignlanguage{english}{and
Cosmic $\hi$ Density $(\OMHI)$ \label{sub:-Mass-Density}}}

\begin{figure}[h]
\begin{lyxlist}{00.00.0000}
\item [{\includegraphics[width=1\columnwidth]{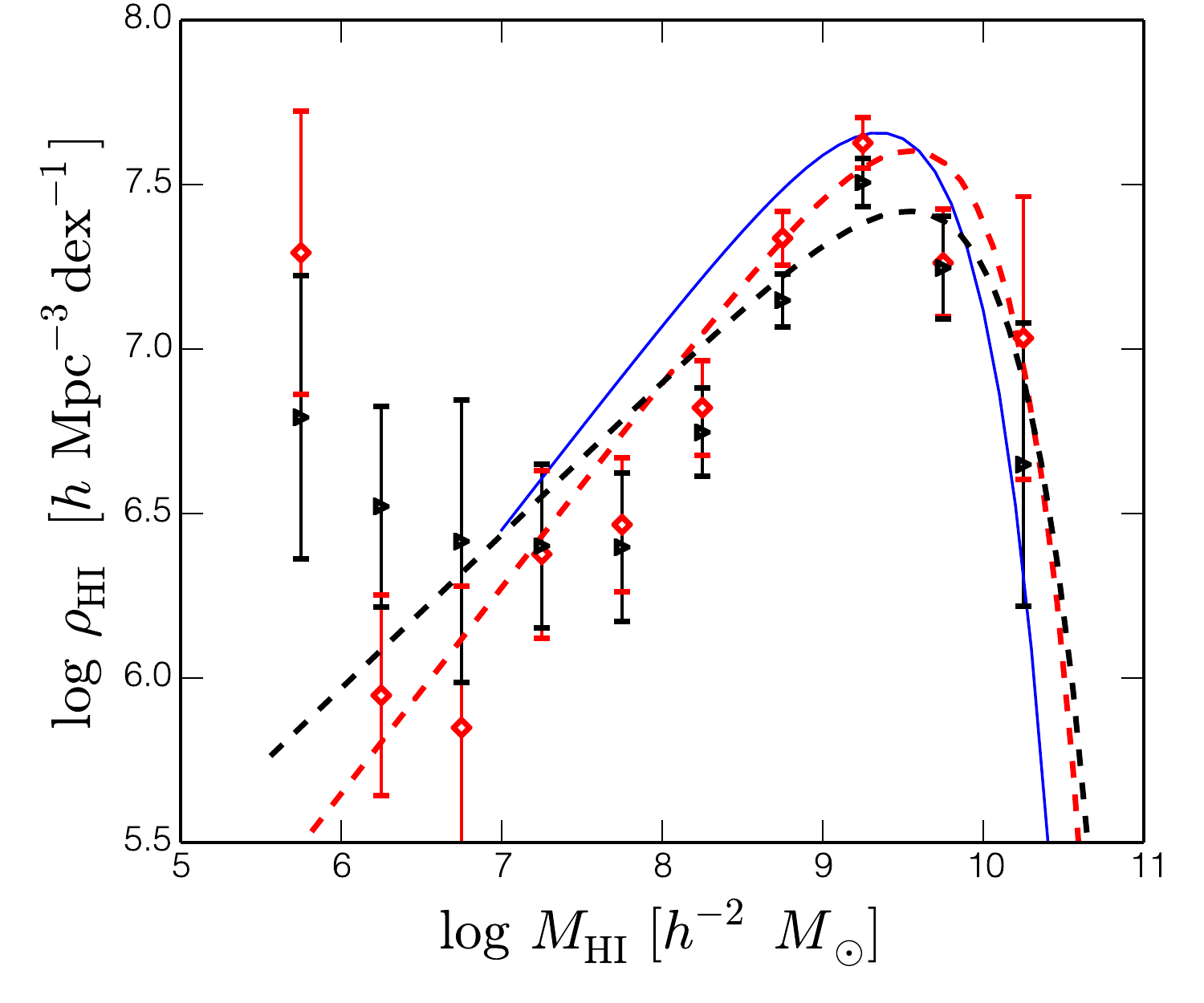}}]~
\end{lyxlist}
\protect\caption[The distribution of the $\hi$ mass density as a function of $\MHI$. ]{The distribution of the $\hi$ mass density as a function of $\MHI$.
The $\hi$ mass density is derived by multiplying the HIMF by the
centre of each $\MHI$ mass bin. The red diamonds and dashed line
show the mass density derived from the $\Sigma1/\Vmaxc$ method and
the black diamonds and dashed line correspond to the $\sum1/\Vmax$
method. The blue line indicates the $\hi$ mass density from HIPASS.
The comparison shows that for AUDS there is a slightly higher contribution
from both faint and bright galaxies to the overall $\hi$ density.\label{fig:rhoF}
}
\end{figure}

Figure~\ref{fig:rhoF} shows the $\hi$ mass density $\rho_{\hi}$
for different $\hi$ masses; we compare the results before and after
cosmic variance correction. The lines indicate the best-fit to $\rho_{\hi}$
using equation~\ref{eq:Schechter-1-1}. The measured slope of the
HIMF has important implication for the contribution to the $\hi$
mass density $\rho_{\mathrm{\hi}}$ of low-mass galaxies. The measured
slope of the AUDS sample that the gas mass density is dominated by
galaxies with masses around $10^{9.7}h^{-2}M_{\odot}$ corresponding
to the knee of the HIMF. Comparing the results of AUDS and HIPASS
(Figure \ref{fig:rhoF}) shows differences between the two surveys.
AUDS detects more galaxies at the low mass end and also detect slightly
more galaxies at the very high mass end.

The total $\hi$~mass density can then be computed by integrating
the best Schechter fit using $\rho_{\hi}=\Gamma(\alpha+2)\times\Phi^{*}\MHI^{*}$,
where $\Gamma$ is the Euler Gamma function and $\alpha,\,\Phi^{*}$
and $\MHI^{*}$ the fit parameters from the Schechter fit. This gives
$\rho_{\hi}=\RI$ for the $\sum\,1/\Vmax$ method and $\rho_{\hi}=\RIC$
for $\sum\,1/\Vmaxc$. We estimated the error using the jackknifing
technique as with the HIMF calculation (Section \ref{sub:-HIMF-Vmax}).

\selectlanguage{english}%
In addition to this, we can also calculate $\rho_{\hi}$ by summing
over individual data points in the $\hi$ density distribution. The
results are $\rho_{\hi}=\RS$ and $\rho_{\hi}=\RSC$ for the uncorrected
and corrected values respectively. The results of the summation and
integration of $\rho_{\hi}$ agree within the $1\sigma$ error-bars
indicating that our survey was able to adequately probe below the
knee of the HIMF to capture most of $\OMHI$.

\selectlanguage{british}%
To compare our results to other measurements we compare the co-moving
$\hi$ density, $\rho_{\hi}$ to the current $(z=0)$ critical density
of the Universe, $\rho_{\mathrm{crit,0}}$, to derive the cosmic $\hi$
density

\begin{equation}
\Omega_{\mathrm{\hi}}=\frac{{\rho_{\mathrm{\hi}}}}{\rho_{\mathrm{{crit}}(z=0)}}=\frac{{8\pi G}}{3H_{0}}\rho_{\hi},\label{eq:OMHeq}
\end{equation}

\selectlanguage{english}%
where G is the Gravitational constant and $H_{0}$ is the Hubble constant
at $z=0$. Note that the \foreignlanguage{british}{definition of $\OMHI$,
consistent with previous work, simply scales the co-moving density
by the current critical density and not by the co-moving, redshift
dependent critical density. }We find $\Omega_{\hi}=(\OS)\times10^{-4}h^{-1}$
before correction and $\Omega_{\hi}=(\OSC)\times10^{-4}h^{-1}$ after
cosmic variance correction, summing up the data points. As noted in
Section \ref{sub:3-2-5-Cosmic-Variance} the cosmic variance correction
has formal uncertainties of 8\% in piggy-backing to the larger 4.2
deg$^{2}$ SDSS field (11\% for a single field, 8\% for two fields)
and 7\% for SDSS as a whole (Driver \& Robotham 2010), giving rise
to a combined systematic uncertainty of 11\%. However, there are additional
factors due to differing bias factors for $\hi\ $ and optical surveys,
and unknown stochasticity factors which may raise this overall uncertainty
(see Chang et al. 2010) which are neglected in this paper.

\selectlanguage{british}%
\begin{figure}
\begin{centering}
\includegraphics[width=0.5\textwidth]{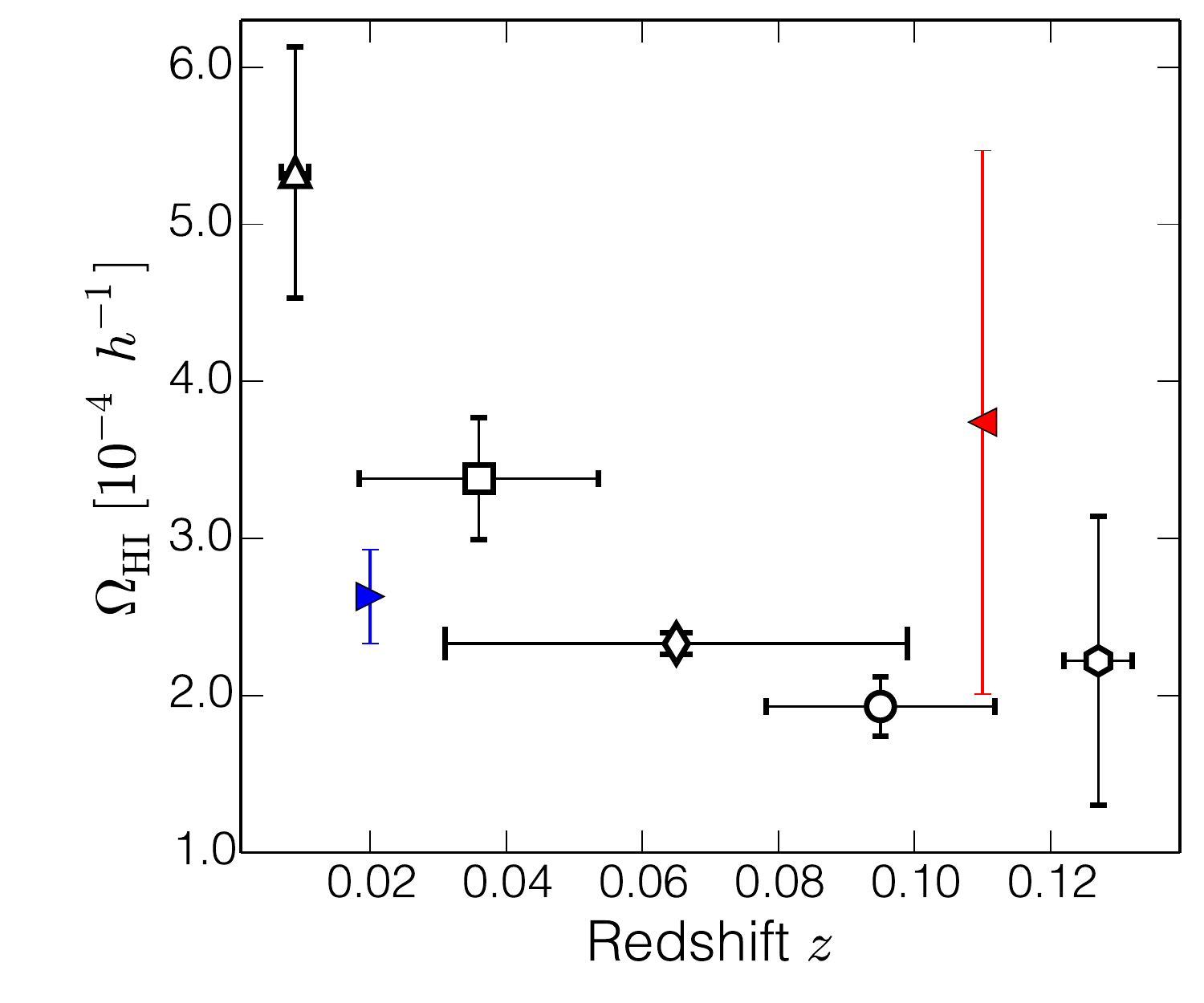}
\par\end{centering}

\protect\protect\caption[Measurements of $\OMHI$ for different redshift bins in AUDS. ]{Measurements of $\OMHI$ for different redshift bins in AUDS (black
points). Note that not all data points are independent from each other
but overlap in redshift (see Table 5). The mean redshift of the subsamples
are $z=0.009$ (triangle), $z=0.036$ (square), $z=0.065$ (diamond),
$z=0.095$ (circle) and $z=0.127$ (hexagon). Comparing the results
to HIPASS (blue right pointing triangle) and the AUDS precursor (red
left pointing triangle) shows little evolution in the measurements
from $z=0$ to $z=0.16$. \label{fig:Evolution-of-the-18} }
\end{figure}

\subsection{Evolution of the Cosmic $\hi$ Density $\OMHI$\label{sub:Evolution-of-the}}

To trace the evolution of cool gas with cosmic time we split our sample
in different bins of redshift. For each sub-sample we derived the
HIMF and calculated $\OMHI$. As the number of the galaxies in the
redshift bins are relatively small we decided to fit the HIMF with
a Schechter function keeping $\MHI^{*}$ and $\alpha$ fixed using
the results we found for the $\sum1/\Vmaxc$ and only fit the normalisation
$\Phi^{*}$. $\OMHI$ is then calculated by integrating over the Schechter
function. First we split the sample in two redshift bins at the mean
redshift of the sample $(\langle z\rangle=0.065)$, creating two samples
with mean redshifts of $\langle z\rangle=0.036$ and $\langle z\rangle=0.095$.
We find $\OMHI=(\OSZlow)\text{\ensuremath{\times}}10^{-4}\, h^{-1}$
for the low redshift sample and $\OMHI=(\OSZhigh)\times10^{-4}\, h^{-1}$
for the high redshift sample. The results indicate a possible decrease
in $\OMHI$ towards the upper end of the redshift range of the sample.

Next we selected the 8 highest redshift galaxies between the redshifts
of 0.119 and 0.132 to probe the high redshift end of our sample. Integrating
over the Schechter function we find ${\normalcolor }\OMHI=(\OSZhighH)\times10^{-4}\, h^{-1}$.
Due to the small numbers of galaxies in this bin Poisson scatter is
the dominant source of error. We also selected a low redshift with
the 8 lowest redshift galaxies. The result is $\OMHI=(\OSZlowL)\times10^{-4}h^{-1}$
agreeing well with results from HIPASS and ALFALFA as well as the
AUDS lower redshift bin.

The results for $\OMHI$ for different redshift bins are summarised
in Figure \ref{fig:Evolution-of-the-18} as well as Table \ref{tab:HIMF-z-tab}.
Note that the measurements of the 8 high and 8 low-redshift galaxies
are not independent of the data binned at the mean redshift of the
sample. It appears likely from this comparison that the low redshift
points are high compared with those at higher redshift. The HIPASS
result agrees better with the high redshift points suggesting that
there is no evolution detected and that the low redshift results may
be subject to cosmic variance errors.

\begin{table}
\protect\caption[Overview over most reliable measurements of $\OMHI$ up to a redshift
of $z=0.2$. ]{Overview of measurements of $\OMHI$ up to a redshift of $z=0.2$.
We calculate a weighted mean of the individual measurement to find
a universal value of $\OMHI$ for this redshift range. \label{tab:OMHI_02}
}

\centering{}%
\begin{tabular}{lccc}
\hline 
Reference  & $\langle z\rangle$  & $\OMHI$  & Method\tabularnewline
 &  & $(10^{-4}h^{-1})$  & \tabularnewline
\hline 
\hline 
\citet{Zwaan2005}  & $0.015$  & $2.6\pm0.3$  & Sources\tabularnewline
\citet{Martin2010}  & $0.025$  & $3.0\pm0.2$  & Sources\tabularnewline
\citet{Freudling2011}  & $0.111$  & $\FreudOMHI$ & Sources\tabularnewline
\citet{delhaize2013}  & $0.028$  & $2.82_{-0.59}^{+0.30}$  & Stacking\tabularnewline
\citet{delhaize2013}  & $0.096$  & $3.19_{-0.59}^{+0.43}$  & Stacking\tabularnewline
\citet{Rhee2013}  & $0.1$  & $2.3\pm0.4$  & Stacking\tabularnewline
\citet{Rhee2013}  & $0.2$  & $2.4\pm0.6$  & Stacking\tabularnewline
\textbf{This paper}  & $0.034$  & $\OSZlow$  & Sources\tabularnewline
\textbf{This paper}  & $0.095$  & $\OSZhigh$  & Sources\tabularnewline
\hline 
Best combined estimate  & $0-0.2$  & $\OSbest$  & \multirow{1}{*}{}\tabularnewline
\hline 
\end{tabular}
\end{table}

\subsection{Discussion}

\begin{figure*}
\begin{centering}
\includegraphics[width=1\textwidth]{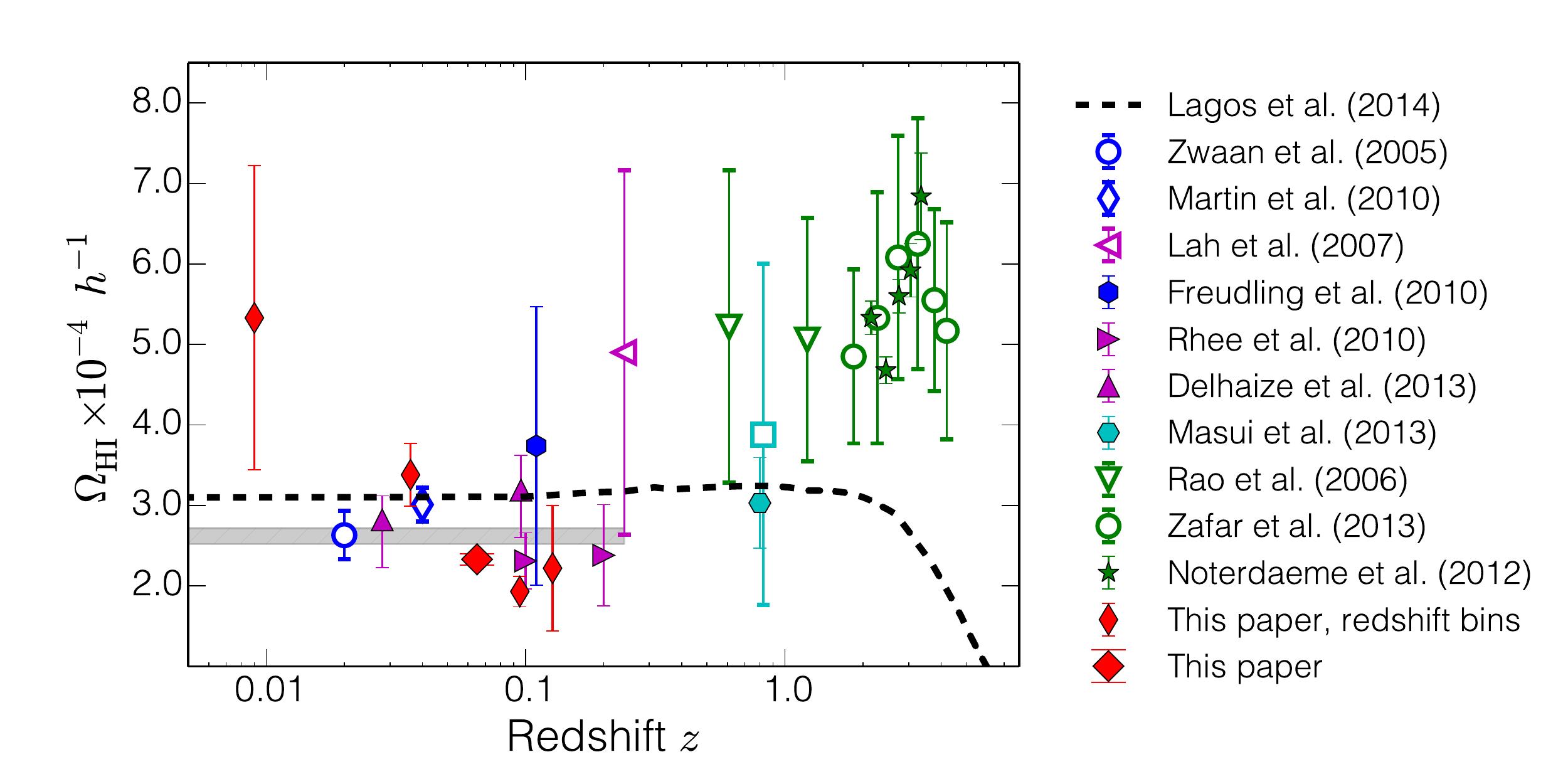}
\par\end{centering}

\selectlanguage{english}%
\protect\protect\caption[Evolution of the cosmic $\OMHI$ density with redshift.]{\selectlanguage{british}%
Evolution of the cosmic $\OMHI$ density with redshift. The colour
corresponds to the type of measurement. Blue: blind 21~cm surveys,
Magenta: $\hi$ stacking, Cyan: Intensity mapping, Green: Measurements
from Ly-$\alpha$ absorption spectra, Red: AUDS. We present the result
using the complete AUDS sample (single thick diamond) as well as the
binned results (thin diamonds). Observations show no significant evolution
in $\OMHI$ out to $z=0.2$. Calculating the best combined estimate
based on all measurements of $\OMHI$ out to this redshift we find
the 1$\sigma$ interval marked in grey. The black dashed line is the
prediction presented by \citet{Lagos2014MNRAS.440..920L}, using semi-analytic
models described in \citet{Lagos2012MNRAS.426.2142L}. \label{fig:The-plot-summarises-19}
\selectlanguage{english}%
}
\selectlanguage{british}%
\end{figure*}

Measuring $\OMHI$ and its evolution with redshift has long been an
important scientific question. 21~cm measurements at low redshift
provide a good constraint for $\OMHI$ in the local Universe. Beyond
that measurements have been more difficult. Until very recently there
has been a huge gap between these local measurements and measurements
at high redshift $z>1.5$ using DLA measurements. Moreover, this period
is marked by a significant change in star formation rate and therefore
interesting for galaxy evolution studies. Successful attempts have
been made using the stacking technique to bridge that intermediate
redshift gap \foreignlanguage{english}{\citep{Rhee2013,delhaize2013}
as well as the intensity mapping technique \citep{Chang2010Natur.466..463C,Masui2013ApJ...763L..20M}
at higher redshift (Figure~\ref{fig:The-plot-summarises-19}). }

\begin{table}
\protect\caption{Results for $\OMHI$ binning the sample in different redshift bins
for AUDS. \label{tab:HIMF-z-tab} }

\centering{}%
\begin{tabular}{ccccc}
\hline 
$\Delta\, z$  & $\langle z\rangle$  & Number  & \selectlanguage{english}%
$\log\,\frac{\MHI}{M_{\odot}}$\selectlanguage{british}%
 & $\OMHI\,$\tabularnewline
 &  &  & $+2\log h$  & \selectlanguage{english}%
$(10^{-4}h^{-1})$\selectlanguage{british}%
\tabularnewline
\hline 
\hline 
$0-0.013$  & $0.009$  & 8  & $6.3-9.3$  & $\OSZlowL^{a}$\tabularnewline
$0-0.065$  & $0.036$  & 52  & $6.3-9.5$  & $\OSZlow^{b}$\tabularnewline
$0.065-0.132$$ $  & $0.095$  & 53  & $8.4-10.3$  & $\OSZhigh^{b}$\tabularnewline
$0.119-0.132$  & $0.127$  & 8  & $8.9-10.3$  & $\OSZhighH^{a}$\tabularnewline
$0-0.132$  & $0.065$  & $\NNo$ & $6.3-10.3$  & $\OSC^{b}$\tabularnewline
\hline 
\multicolumn{5}{l}{$^{a}$ Errors derived using Poisson statistics. }\tabularnewline
\multicolumn{5}{l}{$^{b}$ Errors derived using jackknifing. }\tabularnewline
\end{tabular}
\end{table}

\selectlanguage{english}%
AUDS is the first survey which begins to probe this redshift range
using direct detections (Figure \ref{fig:The-plot-summarises-19}).
The AUDS results combined with the low redshift 21~cm surveys imply
only limited, if any evolutionary effects out to $z=0.2$, corresponding
to a look-back time of $1.7\, h^{-1}\,\mathrm{Gyr}$.

\selectlanguage{british}%
We use these measurements (Table \ref{tab:OMHI_02}) to calculate
a weighted average for $\OMHI$ for $0<z<0.2$. Weighing each measurement
by its error, we find $\OMHI=(\OSbest)\times10^{-4}h^{-1}$. The result
is presented in the Figure \ref{fig:The-plot-summarises-19} as the
grey shaded region indicating the 1~$\sigma$ level. The black dashed
line, also presented in \foreignlanguage{english}{Figure \ref{fig:The-plot-summarises-19}
shows the results from semi-analytic models presented by \citet{Lagos2014MNRAS.440..920L},
using their model as described in \citet{Lagos2012MNRAS.426.2142L}.
They also find a very weak increase in $\OMHI$ over this redshift
range, in agreement with the observational result for $\OMHI$.}

\section{Conclusions\label{sec:3-5-Conclusions}}

\selectlanguage{english}%
In this paper we present early results from AUDS, a deep blind 21~cm
line survey of two selected fields in the redshift range between 0
and 0.16 with a sensitivity of $80\,\mu\mathrm{Jy}$. We detected
a total of $\NNo$ galaxies with masses from \foreignlanguage{british}{$\log(M_{\hi}/M_{\odot})-2\log h=5.6-10.3$.
We used synthetic galaxies to derive a completeness function based
on $\Sint$, $W$ and $\sigma$. We used the SDSS DR7 to correct the
sample for cosmic variance and derived a $\sum1/\Vmaxc$ HIMF which
is well fit by a Schechter function with the parameters: $\alpha=\alC$,
$\Phi^{*}=(\PLC)\,10^{-3}\, h^{3}\,\mathrm{Mpc^{-3}}$ and $\log(M_{\hi}^{*}/M)=(\MiC)+2\log h$,
a result which is in good agreement with results from local surveys
(HIPASS, ALFALFA).}

\selectlanguage{british}%
The co-moving $\hi$ mass density at the mean redshift of the sample
is \foreignlanguage{english}{$\rho_{\hi}=\RSC$ contributing a fraction
$ $$\Omega_{\hi}=(\OSC)\times10^{-4}h^{-1}$ of the critical density
of the Universe. The depth of the survey allows for the first time
the derivation of both the shape and normalization of the HIMF from
a blindly surveyed volume outside of the local Universe. In the volume
that excludes the local Universe, at $z>0.06$, AUDS probes the mass
range from }$\log(M_{\hi}/M_{\odot})-2\log h=8-10.3$\foreignlanguage{english}{.
The derived HIMF is indistinguishable from that derived from local
surveys. These observations constitute strong evidence that the HIMF
did not rapidly evolve in the last billion years.}

At redshifts up to 0.005, AUDS probes the HIMF at masses as low as
$\log(M_{\hi}/M_{\odot})=5.6+2\log h$. We detected twice as many
galaxies as predicted from the local HIMF for $\log(M_{\hi}/M_{\odot})<7+2\log h$.
This might be an indication that the HIMF rises more steeply than
previously thought at the very low mass end. If correct, this finding
implies that the fraction of $\MHI$ contributed by low mass galaxies
may be more significant than previously appreciated.

\section*{Acknowledgements}

We want to thank Ensieh Vaez for help with source finding. LH wants
to thank Jacinta Delhaize and Stefan Westerlund for coding assistance
as well as the ASA for providing travel support. The Arecibo Observatory
is operated by SRI International under a cooperative agreement with
the National Science Foundation (AST-1100968), and in alliance with
Ana G. M�ndez-Universidad Metropolitana, and the Universities Space
Research Association. This research was partially supported by the
ESO DGDF fund. This research made use if the Sloan Digital Sky Survey
archive. The full acknowledgment can be found at http://www.sdss.org.

\end{document}